\documentclass[10pt,a4paper]{article}
\usepackage[T1]{fontenc}
\usepackage[utf8]{inputenc}

\usepackage[english]{babel}
\usepackage{amsmath}
\usepackage{amsfonts}
\usepackage{amssymb}

\usepackage[left=2cm,right=2cm,top=2cm,bottom=2cm]{geometry}

\usepackage{authblk}

\usepackage{hyperref}
\usepackage{graphicx}
\usepackage[dvipsnames]{xcolor}
\newcommand{\mathsfbi}[1]{\boldsymbol{\mathsf{#1}}}
\newcommand{\citep}[1]{(\cite{#1})}

\xdefinecolor{RED}{named}{BrickRed}

\title{Parametric instability and wave turbulence driven by tidal excitation of internal waves}

\author[1]{Thomas Le Reun, Benjamin Favier and Michael Le Bars}
\affil[1]{Aix Marseille Univ, CNRS, Centrale Marseille, IRPHE UMR 7342, Marseille, France}
\date{Accepted for publication in JFM on December 20th 2017}

\begin{document}
\maketitle


\begin{abstract}
We investigate the stability of stratified fluid layers undergoing homogeneous and periodic tidal deformation.
We first introduce a local model which allows to study velocity and buoyancy fluctuations in a Lagrangian domain periodically stretched and sheared by the tidal base flow.
While keeping the key physical ingredients only, such a model is efficient to simulate planetary regimes where tidal amplitudes and dissipation are small. 
With this model, we prove that tidal flows are able to drive parametric subharmonic resonances of internal waves, in a way reminiscent of the elliptical instability in rotating fluids.
The growth rates computed via Direct Numerical Simulations (DNS) are in very good agreement with WKB analysis and Floquet theory.  
We also investigate the turbulence driven by this instability mechanism.
With spatio-temporal analysis, we show that it is a 
weak internal wave turbulence occurring at small Froude and buoyancy Reynolds numbers. 
When the gap between the excitation and the Brunt-V\"ais\"al\"a frequencies is increased, the frequency spectrum of this wave turbulence displays a -2 power law reminiscent of the high-frequency branch of the Garett and Munk spectrum  \citep{garrett_internal_1979} which has been measured in the oceans.
In addition, we find that the mixing efficiency is altered compared to what is computed in the context of DNS of stratified turbulence excited at small Froude and large buoyancy Reynolds numbers and is consistent with a superposition of waves. 
\end{abstract}

\section{Introduction}

Tides affect the whole shape of planetary bodies, from their surface oceans to their liquid cores 
and also their subsurface oceans in the case of icy satellites such as Enceladus \citep{thomas_enceladus_2016,tyler_ocean_2009}. 
Such a mechanical forcing of the topography of these layers is known for driving a wide variety of bulk flows in planetary interiors (see \cite{le_bars_flows_2015} for a review).  
It can directly excite inertial waves \citep{ogilvie_tidal_2004,goodman_dynamical_2009}, drive zonal flows \citep{tilgner_zonal_2007,morize_experimental_2010,sauret_tide-driven_2014,favier_non-linear_2014}, and generate turbulence from various instabilities. 
This is the case for instance of the elliptical instability which, from the combination of the elliptical deformation and the planetary rotation, drives the resonance of inertial waves \citep{kerswell_elliptical_2002}.
Their growth saturates in either sustained turbulence \citep{grannan_tidally_2017,lereun_2017} or cycles of turbulent bursts in which large scale vortices emerge and inhibit the instability \citep{kerswell_secondary_1999,
schaeffer_nonlinear_2010,
barker_non-linear_2013,
lereun_2017}.
Tides can therefore inject energy into planetary fluid layers to sustain intense small-scale motions, this energy being then dissipated via viscous friction. 
This is particularly interesting in the case of planetary cores as it could provide an alternative stirring mechanism to thermal and solutal convection usually invoked to explain dynamo action.  
The large scale tidal flow could also couple with buoyancy effects in the case where these planetary interiors are stratified. 
Many planetary cores are thought to be at least partly stably stratified while maintaining a large scale magnetic field (see \cite{stanley_effects_2008} for a review).
For instance, both experimental and numerical determination of liquid iron's thermal conductivity (see \cite{labrosse_thermal_2015} and references therein) and analysis of the periodicity of the magnetic field fluctuations \citep{buffett_geomagnetic_2014} point towards the existence of a stably stratifed layer at the top of the Earth's core. 
Several routes exist for the excitation of three-dimensionnal turbulent motion within a stably stratified layer by tides or other mechanisms. 
It is a common issue in physical oceanography where the interaction of the large scale tidal flow with ground topography is known for exciting \citep{st._laurent_role_2002}, focusing  \citep{maas_geometric_1995,
bajars_appearance_2013} and scattering \citep{buhler_decay_2011} internal waves which break down into small scale turbulence via triadic resonant  interactions \citep{mackinnon_subtropical_2005,
bourget_experimental_2013,
scolan_nonlinear_2013,
brouzet_energy_2016,brouzet_internal_2017}.
In addition, several studies have strived to examine the resonant excitation of global internal modes by a homogeneous tidal flow without relying on any small-scale topography.
This has been done for instance by \cite{miyazaki_threedimensional_1992,
kerswell_elliptical_1993,
miyazaki_elliptical_1993,
mcwilliams_fluctuation_1998,
aspden_elliptical_2009,guimbard_elliptic_2010} with either radial or vertical stratification compared to deformation, but always in the situation where the Coriolis force is of greater or similar influence compared to the buoyancy effects. 
Whether these resonant instabilities can drive three-dimensionnal turbulence in the bulk of a stratified core or subsurface ocean in a non-rotating case remains to be seen.
In this paper, we derive a local model of fluid planetary interiors, be it a liquid core or a subsurface ocean, which allows to study the idealised limit where stratification completely dominates over rotation, with the stratification axis pointing in any direction relative to that of the tidal deformation. 
We show in particular that tides excite a parametric subharmonic resonance of internal waves. 
Moreover, such an idealised local model allows to thoroughly analyse the turbulent saturation of this tidally-driven resonance.
It is thus shown hereafter that it drives bulk wave turbulence.
Note that parametric resonances of internal waves driven by large-scale homogeneous forcing have already been investigated in particular experimental setups.
\cite{mcewan_parametric_1975} designed a setup to examine how large-scale internal waves spontaneously generate smaller scale oscillations in a stratified tank designed to mimic the avection by large scales. 
In close analogy, \cite{benielli_excitation_1996,benielli_excitation_1998}  showed that vertically shaking a stratified fluid leads to a parametric resonance of internal waves (similarly to the classical Faraday instability) whose growth saturates into turbulence. 
The originality of our study resides in the investigation of the stability of a more realistic homogeneous tidal flow.
We also provide a detailed spatio-temporal analysis of the non-linear break-down into turbulence from the primary resonance. 
This paper is organised as follows. The first part is devoted to introducing the tidal basic flow and developing a local approach to study its stability following the work of \cite{barker_non-linear_2013} and \cite{lereun_2017}.
We carefully introduce buoyancy effects under the Boussinesq approximation.
With this model, we then investigate the resonance of internal waves via direct numerical simulations and Wentzel-Kramers-Brillouin (WKB) analysis of the local model \citep{lifschitz_local_1991}.
We then thoroughly investigate the turbulence resulting from the non-linear saturation of the instability. 
We show that it is best described as a wave turbulence and study the subsequent mixing and dissipation rates.

\section{Local study of the equilibrium base flow} 

\subsection{Tidal base flow}

\begin{figure}
  \centerline{\includegraphics[width=0.5\linewidth]{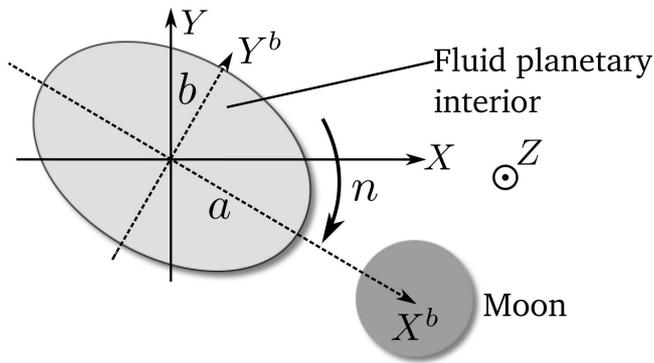}}
  \caption{Schematic representation of a companion orbiting a planet at rate $n$. Tidal interactions induce an ellipsoidal deformation of the whole planet, which is supposed uniform: this corresponds to the behaviour of a fully deformable planet.}
\label{fig:tide_scheme}
\end{figure}

We consider a non-rotating planet undergoing tidal deformation due to a moon orbiting at rate $n \boldsymbol{e}_z$ as pictured in figure \ref{fig:tide_scheme}. 
We suppose the tidal deformation to be ellipsoidal and uniform in the whole fluid planetary interior; this assumption corresponds to planetary cores or oceans in between two boundaries whose response to tidal force is the same as that of the fluid's. 
The other limit, which won't be considered here, is an ocean on top of a non-deformable solid core or inside a non-deformable solid shell for which the ellipticity of the deformation can no longer be considered uniform.
Any planetary fluid layer stands between these two limits. 
In the reference frame following the tidal deformation, the base flow $\boldsymbol{U}_b^{\mathrm{bulge}}$ can be approximated to the following analytical solution \citep{sridhar_tidal_1992,kerswell_elliptical_2002,barker_non-linear_2016,barker_non-linear_2016_2}:
\begin{equation}
\label{eq:basic_flow_B}
\setlength{\arraycolsep}{3pt}
\renewcommand{\arraystretch}{1.0}
\boldsymbol{U}_b^{\mathrm{bulge}} =  \gamma  \left[
\begin{array}{ccc}
0  &  -1 - \beta  &  0    \\
  \displaystyle
   1-\beta &  0  &  0  \\
 0 &  0  &  0  
\end{array}  \right] 
\left[
\begin{array}{c}
  X^{b}     \\
  Y^{b}  \\
  Z^{b}  \\
\end{array}  \right] 
= \mathsfbi{B} \boldsymbol{X}^{b},
\end{equation}
where we have introduced $\gamma = -n$, the rotation rate of the fluid in the orbital frame, by analogy with previous studies \citep{barker_non-linear_2013,lereun_2017}, and $\beta$ the ellipticity of the tidal deformation defined as $\beta = (a^2 - b^2)/(a^2 + b^2)$ (where $a$ and $b$ are the semi-major and semi-minor axes respectively, see figure \ref{fig:tide_scheme}). 
$\boldsymbol{X}^b$ is the position vector whose coordinates are written in the orbital frame rotating with the tidal bulge.
The tidal flow (\ref{eq:basic_flow_B}) is an exact steady solution of theincompressible Navier-Stokes equation in an ellipsoidal planetary fluid layer with stress-free boundary conditions, under the assumption that the tidal potential is small compared to the body self-gravity potential in its unperturbed spherical shape, and neglecting the bulge self-gravity \citep{sridhar_tidal_1992,barker_non-linear_2016,barker_non-linear_2016_2}.
The base flow translates the fact that the streamlines must be lying in the total potential isosurfaces (including gravitational, tidal and rotational contributions) and must match the solid body rotation in the limit $\beta= 0$ measured in the orbital frame. 
They are therefore elliptical and parallel to the $(xOy)$ plane.
In planetary fluid layers, the ellipticity of the tidal flow is usually below $10^{-3}$ \citep{cebron_elliptical_2012} and is about $10^{-7}$ for the Earth's core and oceans (without taking into account any topographical effect).  
In the planetary reference frame the base flow $\boldsymbol{U}_b$ translates, after coordinate change and velocity transformation, into \citep{goodman_local_1993,barker_non-linear_2013}:

\begin{equation}
\label{eq:basic_flow}
\setlength{\arraycolsep}{3pt}
\renewcommand{\arraystretch}{1.0}
\boldsymbol{U}_b = - \gamma \beta \left[
\begin{array}{ccc}
 \sin(2\gamma t)  &  \cos(2 \gamma t)  &  0    \\
  \displaystyle
   \cos(2 \gamma t)  &  -\sin(2\gamma t)  &  0  \\
 0 &  0  &  0  
\end{array}  \right] 
\left[
\begin{array}{c}
  X     \\
  Y  \\
  Z  \\
\end{array}  \right] 
= \mathsfbi{A} (t) \boldsymbol{X}.
\end{equation}
This basic flow is also a solution of the full incompressible Navier-Stokes equations.
Both orbital and planetary frames of reference can be used for simulations.
The orbital frame, in which the base flow is steady and writes as in equation (\ref{eq:basic_flow_B}), is widely used in global numerical simulations of tidal instabilities for it is numerically easier to keep a constant boundary topography (see for instance \cite{grannan_tidally_2017} and \cite{cebron_tidal_2010}). 
Conversely, the planetary frame can be more intuitive: in the case of the Earth's, it corresponds to an observer standing on the solid mantle and feeling nearly two tides per day. 
The base flow in this frame only retains the perturbations induced by tidal deformation.
Note that this basic flow is also a solution of the Navier-Stokes equations in the presence of stable stratification in the Boussinesq approximation. 
Assuming an equilibrium state for which isopycnals are also the surfaces with constant gravitational potential $\phi$, including centrifugal force and tides, the density can be written as a continuous and monotonic function $f$ of $\phi$ so that $\rho \boldsymbol{g} = - f(\phi) \boldsymbol{\nabla} \phi  = -  \boldsymbol{\nabla} F(\phi)$ with $\mathrm{d}F/ \mathrm{d} \phi = f$. 
The equilibrium buoyancy term can then be absorbed in the pressure gradient so that this basic situation is purely barotropic.  
Such a barotropic assumption is valid when the isopycnals can move sufficiently fast to keep track of the orbital motion of the moon and the rotating tidal potential;
it is valid for high Brunt-V\"ais\"al\"a frequency compared to the differential rotation frequency $\gamma$ (see \cite{ogilvie_tidal_2014} for further discussion). 
In the opposite regime where the stratification is weaker, the slow motion of the isopycnals should lead to baroclinicity and excite large scale flow consistently trying to restore the alignment between isopycnal and isopotential.
This paper is rather focused on small scale instabilities in the regime where the Brunt-V\"ais\"al\"a frequency is larger than $\gamma$. 
We therefore discard any baroclinic situation to keep only the global tidal distortion as a source of instability.
Tidally-driven baroclinicity should deserve a study of its own. 
In the following, we introduce buoyancy effects under the Boussinesq approximation. 
We study the tidal instability problem developing a local approach inspired by \cite{barker_non-linear_2013}. 
We use the planetary frame of reference, which is considered to be non-rotating. 
Introducing rotation of this frame at rate $\Omega \boldsymbol{e_z}$ in the model developped hereafter would only require the addition of a Coriolis force in the planetary frame. 
The base flow would not be modified appart from the fluid rotation rate in the orbital reference frame $\gamma$ which would then write $\gamma = \Omega - n$ \citep{barker_non-linear_2013}.
Neglecting planetary rotation is tantamount to assuming the buoyancy effects dominate the dynamics compared to the Coriolis force. 
Studying the interplay between rotation and stratification would require a separate study. 

\subsection{The local approach to the dynamics}
We aim at studying the icompressible perturbations to the basic flow $\boldsymbol{U}_b$ defined in (\ref{eq:basic_flow}). 
Instead of modelling the whole planetary fluid layer, we develop a local model to study this perturbation flow in the neighbourhood of a Lagrangian point $M$ at position $\boldsymbol{X}_0 (t)$ following the elliptical streamlines such that $\dot{\boldsymbol{X}_0} = \boldsymbol{U}_b$. 
This local model will prove particularly convenient to include buoyancy effects as the stratification can be assumed to be locally uniform and linear around the tracked point. 
It has already been proposed in \cite{barker_non-linear_2013} to investigate the elliptical instability in rotating constant density planetary and stellar interiors. 
We reproduce here its derivation to carefully introduce buoyancy effects. 

Let us call $\boldsymbol{v}^{\mathrm{i}}$ the total velocity field in the frame bound to the planet and $\boldsymbol{v}^{\mathrm{c}}$ the total velocity field in the frame bound to $\boldsymbol{X}_0$. $\boldsymbol{v}^{\mathrm{i}}$ satisfies the following Navier-Stokes equation:
\begin{equation}
\label{eq:basic_NS}
\partial_{\tau} \boldsymbol{v}^{\mathrm{i}} + \boldsymbol{v}^{\mathrm{i}} \cdot \boldsymbol{\nabla}_{\boldsymbol{X}} \boldsymbol{v}^{\mathrm{i}}
 ~=~ - \frac{1}{\rho}\boldsymbol{\nabla} P^{\mathrm{i}} +  \nu \boldsymbol{\nabla}_{\boldsymbol{X}}^2 \boldsymbol{v}^{\mathrm{i}}
\end{equation}
where $\tau$ stands for time, $\boldsymbol{\nabla}_{\boldsymbol{X}}$ for the gradient in the $\boldsymbol{X} = (X,Y,Z)$ coordinates, $P^{\mathrm{i}}$ is the pressure and $\rho$ is the density of the fluid. 
$\boldsymbol{v}^{\mathrm{i}}$ is the total velocity and includes the base flow $\boldsymbol{U}_b (\boldsymbol{X})$ and velocity perturbations $\boldsymbol{u}^{\mathrm{i}}$ so that $\boldsymbol{v}^{\mathrm{i}} = \boldsymbol{U}_b (\boldsymbol{X}) + \boldsymbol{u}^{\mathrm{i}}$.
To transform this equation into the frame in translation bound to $\boldsymbol{X}_0$, we process the following coordinate change:
\begin{equation}
\label{eq:coordinates_change}
\left\lbrace 
\begin{array}{rl}
\boldsymbol{x} &= \boldsymbol{X} - \boldsymbol{X}_0 (t) \\
t &= \tau 
\end{array}
\right. ~.
\end{equation} 
The corresponding change in derivatives is $\boldsymbol{\nabla}_{\boldsymbol{X}} = \boldsymbol{\nabla}_{\boldsymbol{x}} =\boldsymbol{\nabla} $ and  $\partial_\tau = \partial_t - \boldsymbol{U}_b \cdot \boldsymbol{\nabla}_{\boldsymbol{x}}$. The velocity measured in the frame bound to $\boldsymbol{X}_0$ is $\boldsymbol{v}^{\mathrm{c}} = \boldsymbol{v}^{\mathrm{i}} - \boldsymbol{U}_b (\boldsymbol{X}_0 ,t)$. 
Transforming the equation (\ref{eq:basic_NS}) into this frame yields:
\begin{equation}
\label{eq:X0_NS}
~\partial_t \boldsymbol{v}^{\mathrm{c}} + \partial_t \boldsymbol{U}_b (\boldsymbol{X}_0 )
+ \boldsymbol{v}^{\mathrm{c}} \cdot \boldsymbol{\nabla} \boldsymbol{v}^{\mathrm{c}}
~ 
= 
- \frac{1}{\rho}\boldsymbol{\nabla} P^{\mathrm{i}} (\boldsymbol{X}_0 + \boldsymbol{x} ) +  \nu \boldsymbol{\nabla}^2  \boldsymbol{v}^{\mathrm{c}} ~.
\end{equation}
The acceleration term $\partial_t \boldsymbol{U}_b (\boldsymbol{X}_0 ) $ is regarded as a volume force. 
In the frame bound to $\boldsymbol{X}_0$ the Navier-Stokes equation reads:
\begin{equation}
\partial_t \boldsymbol{v}^{\mathrm{c}} 
+ \boldsymbol{v}^{\mathrm{c}} \cdot \boldsymbol{\nabla} \boldsymbol{v}^{\mathrm{c}} 
~=~ 
- \frac{1}{\rho} \boldsymbol{\nabla} P^{c} 
-   \partial_t  \boldsymbol{U}_b (\boldsymbol{X}_0 )
+  \nu \boldsymbol{\nabla}^2  \boldsymbol{v}^{\mathrm{c}} ~
\end{equation}
where we have introduced $P^c (\boldsymbol{x})  = P^{\mathrm{i}} (\boldsymbol{X}_0 + \boldsymbol{x} ) $.
As $\boldsymbol{v}^c = \boldsymbol{v}^{\rm{i}} - \boldsymbol{U}_b (\boldsymbol{X}_0)$, it is straightforward that $\boldsymbol{v}^c = \mathsfbi{A} \boldsymbol{x} + \boldsymbol{u}^c$ with $\boldsymbol{u}^c = \boldsymbol{u}^{\rm{i}} = \boldsymbol{u} $.
In the neighbourhood of $\boldsymbol{X}_0$, the perturbed flow satisfies the following equation:
\begin{equation}
\label{eq:NS_perturbed}
 \partial_t \boldsymbol{u} + \mathsfbi{A} (t) \boldsymbol{x} \cdot \boldsymbol{\nabla}\boldsymbol{u} 
+ \mathsfbi{A}(t) \boldsymbol{u} 
+ \boldsymbol{u} \cdot \boldsymbol{\nabla } \boldsymbol{u}
 = -  \frac{1}{\rho} \boldsymbol{\nabla} P^c  
 -   \partial_t \boldsymbol{U}_b (\boldsymbol{X}_0 ) + \nu \boldsymbol{\nabla}^2 
\boldsymbol{u}~
\end{equation}
along with the incompressibility condition $\boldsymbol{\nabla} \cdot \boldsymbol{u} = 0$.

\subsection{Lagragian effects of the base flow}

\begin{figure}
\centering
\includegraphics[width=\linewidth]{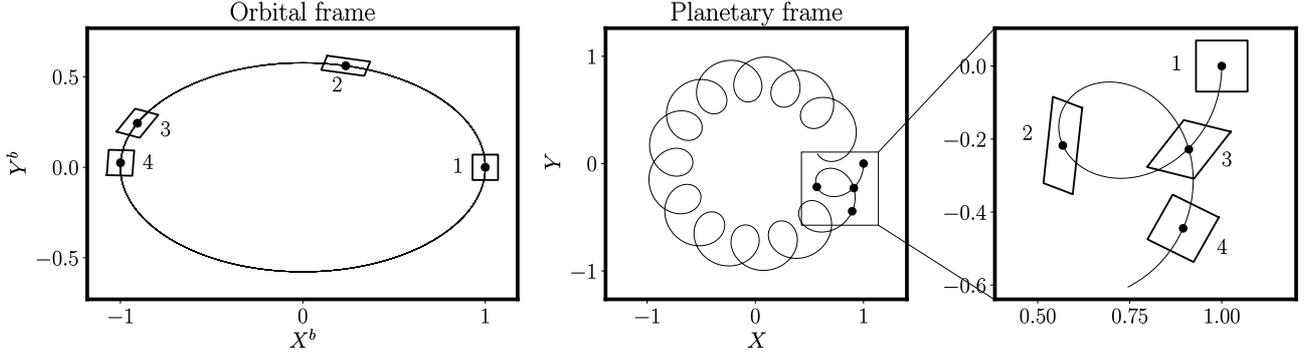}
\caption{
Trajectory of an initially square patch of fluid for an ellipticity $\beta= 0.5$ represented in the orbital frame (left) in which the bulge has a stationnary shape, and in the planetary frame (right). The four configurations correspond to the same instants. The solid black line materialises the trajectory of the center of the patch. This picture highlights the periodic stretching and shearing undergone by the lagrangian parcel. 
}
\label{fig:patch_trajectory}
\end{figure}

This paragraph aims at exhibiting the Lagrangian trajectory of the point $M$ at $\boldsymbol{X}_0$ to provide a better understanding of the model derived in the preceding paragraphs. 
The Lagrangian equation $\dot{\boldsymbol{X}_0} = \boldsymbol{U}_b (\boldsymbol{X}_0)$ can be solved analytically and the position of $M$ at any time can be related to the initial position $\left( X_{0i}, Y_{0i},Z_{0i} \right)$ by the following relation:
\begin{equation}
\label{eq:lagragian_trajectory}
\setlength{\arraycolsep}{2pt}
\renewcommand{\arraystretch}{1.}
\boldsymbol{X}_0= \mathrm{Rot}(-\gamma t) \left[
\begin{array}{ccc}
 \cos(\varpi t)  &  -\mathcal{E} \sin(\varpi t)  &  0    \\
  \displaystyle
   \frac{1}{\mathcal{E}} \sin(\varpi t)  &  \cos(\varpi t)  &  0  \\
 0 &  0  &  1  
\end{array}  \right] 
\left[
\begin{array}{c}
  X_{0i}     \\
  Y_{0i} \\
  Z_{0i}  \\
\end{array}  \right] 
\end{equation}
where $\mathrm{Rot}(-\gamma t)$ is a rotation matrix of angle $-\gamma t$ around the $Z$ axis. 
$\varpi$ and $\mathcal{E}$ are defined as follows:
\begin{equation}
\label{eq:def_omega_E}
\varpi = \gamma \sqrt{1-\beta^2} ~~~~~\mbox{and}~~~~~
\displaystyle  \mathcal{E} = \sqrt{\frac{1+\beta}{1- \beta}} ~~.
\end{equation}
The corresponding trajectories are plotted in figure \ref{fig:patch_trajectory} for $\beta= 0.5$, $\gamma= 1$ and an initial condition $(1,0,0)$. 
The equivalent trajectory in the orbital frame (consistently aligned with the moon) is also indicated for comparison.
In the small $\beta$ limit, it can be shown that, in the planetary frame, the Lagrangian particle rotates around the $Z$ axis at rate $-\gamma \beta^2 /2$, the corresponding acceleration being negligible. 
Around this slow mean rotation corresponding to a Stokes drift with velocity $\sim \gamma \beta^2 \ell$ with $\ell$ the average distance from the centre of the planetary body, the particle also accomplishes epicycles at a much higher rate $\gamma/2$ with a displacement of order $\beta \ell$ and velocity $\beta \gamma \ell$. 
To materialise the local effects of the basic flow, we also plot in figure \ref{fig:patch_trajectory} the trajectories of four points forming a square pattern around the tracked point. 
It can be noticed that this pattern is stretched and sheared during an epicycle and rotates as the particle moves around the $Z$ axis. 
Note that the slow mean rotation around the $Z$ axis is exaggerated in figure \ref{fig:patch_trajectory} because of the very high ellipticity.
This periodic stretching and shearing may drive internal wave parametric instability, in close analogy to the elliptical instability in rotating fluids \citep{kerswell_elliptical_2002}. 

\subsection{Introducing buoyancy in the local Navier-Stokes equation}
The local model introduced above is particularly useful when introducing buoyancy effects, be it due to temperature or solute concentration. 
In a sufficiently small patch, the background stratification can be assumed to be uniform, while being mostly radial in global planetary layers. 
Let us call $\boldsymbol{S}(t)$ the background active scalar gradient such that the total scalar field $T$ can be written as:
\begin{equation}
\label{eq:temperature_definition}
T = T_0 + \boldsymbol{S}(t) \cdot \boldsymbol{x} + \vartheta
\end{equation}
where $\vartheta$ is the scalar fluctuation around the background stratification. 
As it will be pointed out later, it is required to include a temporal dependence in the stratification to account for periodic stretching induced by the background base flow defined in (\ref{eq:basic_flow}) and illustrated in figure \ref{fig:patch_trajectory}.  
The buoyancy effects are first introduced in equation (\ref{eq:NS_perturbed}), in the Boussinesq approximation, via the volume effective gravitational force $\rho_0 ( \boldsymbol{g} -  \partial_t \boldsymbol{U}_b (\boldsymbol{X}_0 )) ( 1 - \alpha (T - T_0) )$ where $\alpha$ is an isobaric thermal expansion coefficient. 
As we assume the basic state to be barotropic, the term $\rho_0 ( \boldsymbol{g} -  \partial_t \boldsymbol{U}_b (\boldsymbol{X}_0 )) ( 1 - \alpha \boldsymbol{S} \cdot \boldsymbol{x} ) $ can be absorbed in a modified pressure $\Pi$. 
Including buoyancy effects, the equations (\ref{eq:NS_perturbed}) now reads:
\begin{equation}
\label{eq:NS_buoy_perturbed}
\partial_t \boldsymbol{u} 
+ \mathsfbi{A} (t) \boldsymbol{x} \cdot \boldsymbol{\nabla}\boldsymbol{u} 
+ \mathsfbi{A}(t) \boldsymbol{u} 
+ \boldsymbol{u} \cdot \boldsymbol{\nabla } \boldsymbol{u}   
= - \boldsymbol{\nabla} \Pi 
- \alpha \left[~~\boldsymbol{g} 
-  \partial_t   \boldsymbol{U}_b (\boldsymbol{X}_0 ) ~~\right]  \vartheta +  \nu \boldsymbol{\nabla}^2 
\boldsymbol{u}~.
\end{equation}
Note that the field $\boldsymbol{u}$ also satisfies the incompressibility condition $\boldsymbol{\nabla} \cdot \boldsymbol{u} = 0$. 
Two possible instability sources are worth considering. 
In the RHS of (\ref{eq:NS_buoy_perturbed}), the Lagrangian advection of the studied patch translates into an effective gravity with varying intensity. 
Such a forcing has already been shown to parametrically excite internal waves with a growth rate proportional to the oscillating acceleration amplitude \citep{mcewan_parametric_1975,benielli_excitation_1998}.
The base flow is also coupled in the left hand side to the velocity perturbation; in the analogue context of rotating flows, the induced tidal stretching and shearing is well known for triggering parametric excitation of a pair of inertial waves. 
This instability mechanism has also been studied in the context of strained vortices with a stratification aligned with the background vorticity \citep{miyazaki_threedimensional_1992,miyazaki_elliptical_1993,
kerswell_elliptical_1993,
mcwilliams_fluctuation_1998,
aspden_elliptical_2009,
guimbard_elliptic_2010}.
Whether a similar stirring mechanism occurs in purely stratified fluids, \textit{i.e.} with no background vorticity, has never been investigated to the best of our knowledge.  
In this paper, we propose to drop the gravity-driven parametric excitation to focus on tidal stretching and shearing effects. 
To support this approximation, we suggest to compare the order of magnitude of the expected growth rate of both instabilities. 
In the case of excitation by gravity variations, the amplitude of the forcing is  proportional to the acceleration of a Lagragian particle during an epicycle: $\beta \ell \gamma^2$. 
Following \cite{benielli_excitation_1998}, the corresponding growth rate $\sigma_g$ should scale like $\sigma_g \sim \gamma (\beta \ell \gamma^2)/g$ where $g$ is the mean intensity of gravity. 
On the other hand, if the coupling between the base flow and the velocity perturbation acts as in the elliptical instability in rotating fluids, the growth rate $\sigma_e$ should then scale like $\sigma_e \sim \beta \gamma$ (see eq. (4) in \cite{barker_non-linear_2013}). 
As a consequence, $\sigma_e/\sigma_g \sim g / (\ell \gamma^2) $. 
As $\gamma$ is at most comparable to the spin rate of the considered planetary body, this ratio equivalently compares the self-gravity of the body to the centrifugal acceleration.
It should then always be large to ensure self-cohesion. 
For instance, for the Moon-Earth system, $\gamma = 2 \pi /(1 \mbox{day})$; at the Core-Mantle boundary $\ell \sim 3 \times 10^{3}$ km and $g \sim 10$ m.s$^{-2}$. 
The ratio $\sigma_e/\sigma_g$ is about $3 \times 10^2$ at the Core-Mantle boundary and $1.5 \times 10^{2}$ at the surface of the Earth, which justifies dropping the varying gravity forcing term. 
Note however that in the case of a confined layer above a non-deformable core, the elliptical base flow, which is then no longer described by (\ref{eq:basic_flow}), would create large-amplitude lateral flows whose contribution to the effective gravity would not necessarily be negligible. 
Although of interest for instance for the Earth's oceans, we do not consider the latter case here to focus on a fully deformable object. 
\subsection{The buoyancy equation and time dependence of stratification}\label{buoyancy_equation}
\begin{figure}
	\centering
  \includegraphics[height=0.24\linewidth]{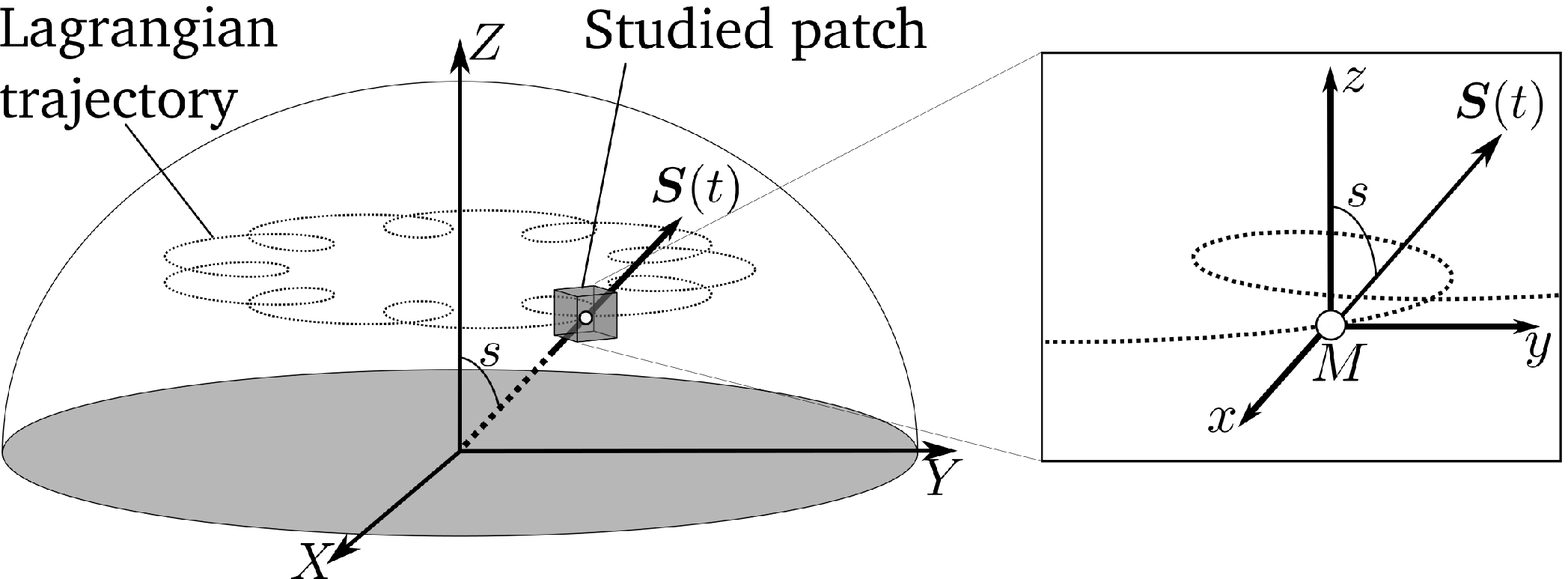} 
  \includegraphics[height=0.24\linewidth]{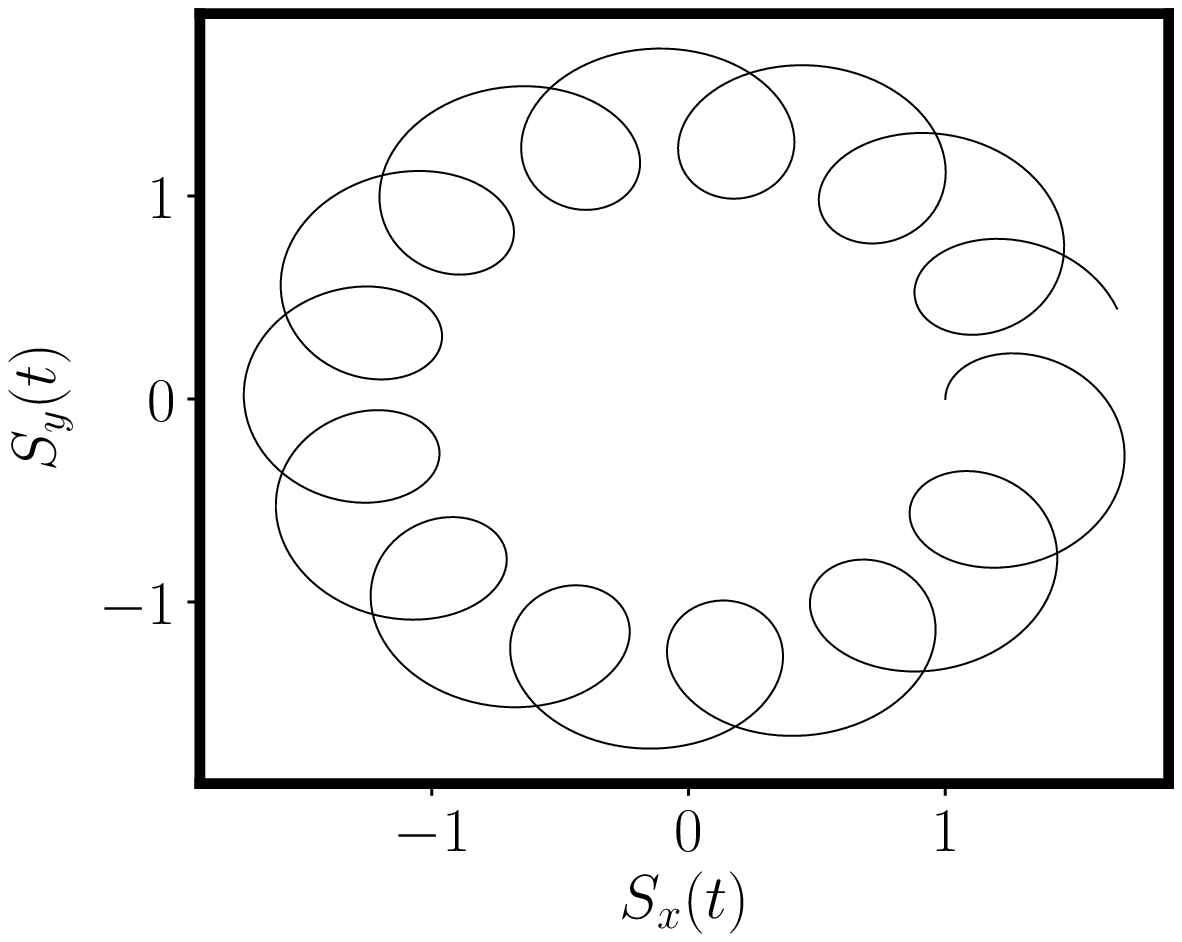}
  \caption{ \textbf{Left:} schematic picture linking stratification seen from a global and a local view. The angle $s$ between stratification and $(OZ)$ is then a proxy for the latitude at which the patch is located. Note that the local axes $(Mxyz)$ are in translation around the $Z$ axis. \textbf{Right:} time evolution of the $x$ and $y$ components of the stratification. Note that the slow rotation of $\boldsymbol{S}$ is due to the translative motion of the Lagrangian particle around $Z$.  }
\label{fig:stratif_time}
\end{figure}

The Navier-Stokes equation (\ref{eq:NS_buoy_perturbed}) is coupled to the advection-diffusion equation for the scalar $T$:
\begin{equation}
\label{eq:transport_T}
\partial_t T + ( \mathsfbi{A}(t) \boldsymbol{x} + \boldsymbol{u} ) \cdot \boldsymbol{\nabla} T = \kappa \nabla^2 T
\end{equation}
where $\kappa$ is a diffusivity coefficient assumed to be constant.
Plugging the assumption (\ref{eq:temperature_definition}) in (\ref{eq:transport_T}) leads to the following modified advection-diffusion equation:
\begin{equation}
\label{eq:modified_transport}
\frac{\mathrm{d} \boldsymbol{S}}{\mathrm{d} t} \cdot \boldsymbol{x} 
+ \frac{\partial \vartheta}{\partial t}
+ \mathsfbi{A}(t)\boldsymbol{x} \cdot \boldsymbol{S} +\mathsfbi{A}(t)\boldsymbol{x} \cdot \boldsymbol{\nabla} \vartheta 
+ \boldsymbol{S} \cdot \boldsymbol{u}
+ \boldsymbol{u} \cdot \boldsymbol{\nabla} \vartheta = 
 \kappa \nabla^2 \vartheta ~ .
\end{equation}
Assuming that in the equilibrium state there is no perturbation, \textit{i.e.} 
$(\boldsymbol{u},\vartheta) = (\boldsymbol{0},0)$,  compels the time evolution of the stratification vector $\boldsymbol{S}(t)$ to follow:
\begin{equation}
\label{eq:stratif_time_evolution}
\frac{\mathrm{d} \boldsymbol{S}}{\mathrm{d} t} = - \mathsfbi{A}^{T} (t) \boldsymbol{S}  
\end{equation}
where $\mathsfbi{A}^{T}$ stands for the transpose of $\mathsfbi{A}$ given in equation (\ref{eq:basic_flow}). 
The periodic stretching and shearing induced by the base flow, as represented in figure \ref{fig:patch_trajectory}, impacts the local background density profile.
The equation (\ref{eq:stratif_time_evolution}) can be solved analytically to obtain that an initial stratification $\boldsymbol{S}_0 = (S_{0x}, S_{0y}, S_{0z})$ evolves in the following way:  
\begin{equation}
\label{eq:stratification_time}
\setlength{\arraycolsep}{2pt}
\renewcommand{\arraystretch}{1.}
\boldsymbol{S} (t)= \mathrm{Rot}(-\gamma t) \left[
\begin{array}{ccc}
 \cos(\varpi t)  &  \displaystyle - \frac{1}{\mathcal{E}} \sin(\varpi t)  &  0    \\
  
  \mathcal{E} \sin(\varpi t)  &  \cos(\varpi t)  &  0  \\
 0 &  0  &  1  
\end{array}  \right] 
\left[
\begin{array}{c}
  S_{x0}     \\
  S_{y0}  \\
  S_{z0}  \\
\end{array}  \right]  = \mathrm{Rot}(-\gamma t) \hat{\mathsfbi{R}} (t) \boldsymbol{S}_0
\end{equation}
where $\mathcal{E}$, $\mathrm{Rot}(\gamma t)$ and $\varpi$ are the same as defined in (\ref{eq:lagragian_trajectory}). 
A typical time evolution of $\boldsymbol{S}(t)$ is pictured in figure \ref{fig:stratif_time}. 
The initial stratification can arbitrarily be set in the $(xMz)$ plane. 
It is then fully parameterised by the angle $s$ such that $\boldsymbol{S}_0 = S_0 ( \sin s, 0 ,\cos s ) = S_0 \tilde{\boldsymbol{s}}_0 $; it represents the mean latitude at which the tracked patch is located. 

Note that to avoid the spontaneous appearance of baroclinic instability, the gravity has to change its direction to stay aligned with the buoyancy gradient. 
This is consistent with the fact that the point $\boldsymbol{X}_0$ is following an elliptical streamline included in an equipotential surface of the total gravitational field including tidal force and  self-gravitation of the unperturbed state; the gravitational field must remain perpendicular to streamlines.  
We arbitrarily choose to keep the gravitational field amplitude constant throughout time. 
Calling $g$ the gravity intensity, the final set of local equations is, in addition to (\ref{eq:stratif_time_evolution}):
\begin{eqnarray}
\left\lbrace
\begin{array}{rl}
\displaystyle
\frac{\partial \boldsymbol{u}}{\partial t} + \mathsfbi{A} (t) \boldsymbol{x} \cdot \boldsymbol{\nabla}\boldsymbol{u} + \mathsfbi{A}(t) \boldsymbol{u} + \boldsymbol{u} \cdot \boldsymbol{\nabla } \boldsymbol{u} 
& \displaystyle = - \boldsymbol{\nabla} \Pi + \alpha g \frac{\boldsymbol{S}}{ \Vert \boldsymbol{S} \Vert} \vartheta +  \nu \boldsymbol{\nabla}^2 
\boldsymbol{u}~ \label{eq:final_dim_u} \\
\displaystyle
\frac{\partial \vartheta}{\partial t}
 +\mathsfbi{A}(t)\boldsymbol{x} \cdot \boldsymbol{\nabla} \vartheta 
+ \boldsymbol{u} \cdot \boldsymbol{\nabla} \vartheta 
&= - \boldsymbol{S}(t) \cdot \boldsymbol{u}
+ \kappa \nabla^2 \vartheta ~ .
\end{array} \right. \label{eq:final_dim_theta}
\end{eqnarray}
%
%

\subsection{Conclusion on the equations and the local model}
We will consider hereafter a patch of typical size $L$. Typical time and velocity scales are then given by $ 1/\gamma $ and $L \gamma$. 
The initial stratification amplitude $S_0$ can be used to build a buoyancy scale $L S_0$. 
With those definitions, the dimensionless velocity-temperature dynamics satisfies:
\begin{equation}
\label{eq:NS_buoy_adim}
\left\lbrace
\begin{array}{rl}
\displaystyle
\frac{\partial \boldsymbol{u}}{\partial t} + \mathsfbi{A} (t) \boldsymbol{x} \cdot \boldsymbol{\nabla}\boldsymbol{u} + \mathsfbi{A}(t) \boldsymbol{u} + \boldsymbol{u} \cdot \boldsymbol{\nabla } \boldsymbol{u}
& \displaystyle = - \boldsymbol{\nabla} \Pi  + N^2 \vartheta \boldsymbol{e}_s +  \frac{1}{Re} \boldsymbol{\nabla}^2 
\boldsymbol{u}~  \\
\displaystyle
\frac{\partial \vartheta}{\partial t}
 +\mathsfbi{A}(t)\boldsymbol{x} \cdot \boldsymbol{\nabla} \vartheta 
+ \boldsymbol{u} \cdot \boldsymbol{\nabla} \vartheta 
&=  \displaystyle - \boldsymbol{\tilde{S}}(t) \cdot \boldsymbol{u}
+  \frac{1}{Pr ~Re} \nabla^2 \vartheta ~ \\
\boldsymbol{\nabla} \cdot \boldsymbol{u} & = 0 
\end{array} \right.  
\end{equation}
where we have introduced the dimensionless Brunt-V\"ais\"al\"a frequency $N$ such that $N^2 = \alpha g S_0 /\gamma$. $Re$ is  the Reynolds number $L^2 \gamma/\nu$ and $Pr$ is the Prandtl number $\nu/\kappa$. 
$\boldsymbol{\tilde{S}}(t)$ and $\boldsymbol{e}_s$ are defined as follows:
\begin{equation}
\label{eq:stratif_adim}
\boldsymbol{\tilde{S}}(t) = \mathrm{Rot}(-\gamma t) \hat{\mathsfbi{R}} (t) \boldsymbol{S}_0 / S_0 = \mathrm{Rot}(-\gamma t) \hat{\mathsfbi{R}} (t) \tilde{\boldsymbol{s}}_0  ~~~~\mbox{and} ~~~~ \boldsymbol{e}_s = \boldsymbol{\tilde{S}}(t) ~/ \parallel \boldsymbol{\tilde{S}}(t) \parallel
\end{equation}

This set of equations is particularly convenient as it reduces a global problem with a non-trivial ellipsoidal geometry to a local one in Cartesian coordinates with uniform stratification. 
It retains all the key ingredients to understand the homogeneous dynamics of tidally-forced flows in stratified layers while avoiding to account for boundary layers. 

\subsection{The local model in the orbital frame}
\label{orbital_frame}
The same analysis can be performed in the orbital frame tracking the elliptical deformation rotation at rate $n \boldsymbol{e}_z$.
The set of equations obtained is very similar to (\ref{eq:NS_buoy_adim}) except that $\mathsfbi{A}(t)$ must be replaced by the matrix $\mathsfbi{B}$ defined in (\ref{eq:basic_flow_B}) and a Coriolis acceleration $2 n \boldsymbol{e}_z \times \boldsymbol{u}$ must be added to the left hand side of the momentum conservation. 
In addition, the time evolution of the stratification vector reads $\boldsymbol{\tilde{S}}(t)=  \hat{\mathsfbi{R}} (t) \tilde{\boldsymbol{s}}_0 $ where $ \hat{\mathsfbi{R}}$ has been introduced in (\ref{eq:stratification_time}).
The two frames of reference are equivalent; nevertheless we prefer the planetary frame as it allows to introduce planetary rotation with the mere intuitive addition of a Coriolis acceleration.
In the non stratified case, this frame allows to clearly identify the inertial waves frequencies \citep{barker_non-linear_2013,lereun_2017}. 
We have therefore considered the model in this frame to facilitate future works concerned with the interplay between rotation and stratification.  

\subsection{Direct numerical simulations in a shearing box}\label{spectral_equations}
The dynamics of the perturbations $\boldsymbol{u}$ to the equilibrium state can be simulated via a decomposition of $\left\lbrace \boldsymbol{u} ,\Pi, \vartheta \right\rbrace$ into Kelvin modes such that:
\begin{equation}
\label{def_spectral}
\left\lbrace \boldsymbol{u} ,\Pi, \vartheta \right\rbrace 
~ = ~ \sum_{\boldsymbol{k}} 
\left\lbrace \hat{\boldsymbol{u}}_{\boldsymbol{k}}(t),\hat{\Pi}_{\boldsymbol{k}}(t),\hat{\vartheta}_{\boldsymbol{k}}(t) \right\rbrace 
~ e^{ i \boldsymbol{k}(t) . \boldsymbol{x}} ~.
\end{equation}
In close analogy to the model developed in section \ref{buoyancy_equation} where stratification is found to be time dependent, evolving the wave vectors $\boldsymbol{k}$ accounts for the periodic shearing induced by the base flow $\boldsymbol{U}_b$.
For all $\boldsymbol{k}$, the equations (\ref{eq:NS_buoy_adim}) are equivalent to: 
\begin{equation}
\label{eq:full_SNOOPY}
\renewcommand{\arraystretch}{2.0}
\left\lbrace
\begin{array}{rl}
\displaystyle \frac{\mathrm{d} \boldsymbol{k} }{\mathrm{d} t}~ &= - ~  \mathsfbi{A}^T (t)~ \boldsymbol{k} = - ~  \mathsfbi{A} (t)~ \boldsymbol{k}  \\
\displaystyle \frac{\mathrm{d} \hat{\boldsymbol{u}}_{\boldsymbol{k}} }{\mathrm{d} t} 
~&=~ 
\displaystyle - \mathsfbi{A}(t) \hat{\boldsymbol{u}}_{\boldsymbol{k}}  
 - i \boldsymbol{k} \hat{\Pi}_{\boldsymbol{k}} 
 + N^2 \hat{\vartheta}_{\boldsymbol{k}} \boldsymbol{e}_s 
 - \frac{k^2}{Re} \hat{\boldsymbol{u}}_{\boldsymbol{k}} 
 - \widehat{(\boldsymbol{u} \cdot \boldsymbol{\nabla} \boldsymbol{u} )}_{\boldsymbol{k}} \\
\displaystyle \frac{\mathrm{d} \hat{\vartheta}_{\boldsymbol{k}} }{\mathrm{d} t} 
~&=~ \displaystyle - \boldsymbol{\tilde{S}}(t) \cdot \hat{\boldsymbol{u}}_{\boldsymbol{k}} 
- \frac{k^2}{Re ~ Pr} \hat{\boldsymbol{u}}_{\boldsymbol{k}}
- \widehat{(\boldsymbol{u} \cdot \boldsymbol{\nabla} \vartheta)}_{\boldsymbol{k}}
\end{array}
\right. .
\end{equation}
A similar derivation was first carried out by \cite{barker_non-linear_2013} to study the tidally-driven elliptical instability in rotating non-stratified fluids. 
Note that the time evolution of $\boldsymbol{k} (t)$ is the same as that of the stratification (\ref{eq:stratif_time_evolution}).
This  shearing box development allows to simulate the perturbations to the base flow with efficient pseudo-spectral methods, as originaly devised by \cite{rogallo_numerical_1981},  for instance implemented in the \textsc{Snoopy} code initiated by \cite{lesur_relevance_2005} that we use here.
Apart from the time evolution of the wave vectors which is known analytically, the code solves the ODEs (\ref{eq:full_SNOOPY}) with a fourth order Runge-Kutta method and applying a 2/3-rule for dealiasing the non-linear terms.

\section{Stability analysis}
\subsection{Direct Numerical Simulations}
We first investigate the stability of the elliptical base flow by performing direct numerical simulations of the full problem, including viscosity and non-linearities. 
This is done with the \textsc{Snoopy} code mentioned in section \ref{spectral_equations} which solves the equations (\ref{eq:full_SNOOPY}).
The Reynolds number $Re$ is usually set between $10^6$ and $10^7$ while the Prandtl number $Pr$ is kept constant at $1$.  
The resolution used is up to $96$ modes in each direction in a square box of size $L$ which is used as a length scale.
The simulations are initiated by a broad-band noise with $k/(2\pi)$ ranging from $4$ to $20$. 
The time evolution of the volume-averaged kinetic energy of the fluctuations is tracked until an exponential phase is reached from which a growth rate is derived. 

\subsection{WKB and Floquet analysis}
Along with solving the full problem, we examine the linear inviscid limit of equations (\ref{eq:NS_buoy_adim}) via a Wentzel-Kramers-Brillouin analysis \citep{lifschitz_local_1991}. 
It is easier to perform the stability analysis in the orbital frame, following the rotation of the elliptical deformation, where the base flow matrix $\mathsfbi{B}$ does not depend on time (see paragraph \ref{orbital_frame} for the corresponding change in equations (\ref{eq:NS_buoy_adim})) . 
We then assume that the velocity, pressure and buoyancy fluctuations around the equilibrium state may be written as follows:
\begin{equation}
\label{eq:WKB_assumption}
\left\lbrace \boldsymbol{u} ,\Pi, \vartheta \right\rbrace = \left\lbrace \boldsymbol{a} ,p , \Theta \right\rbrace e^{i\displaystyle \frac{\phi(\boldsymbol{x},t)}{\eta} }
\end{equation}
where $\eta$ is a small parameter accounting for the quick wave-like spatial variations of the phase $\phi$ compared to the secular evolution of amplitudes $\boldsymbol{a}$, $p$ and $\Theta$. 
Plugging the assumption (\ref{eq:WKB_assumption}) into (\ref{eq:NS_buoy_adim}) in the linear inviscid limit and performing a series expansion in $\eta$ lead to the following set of equations \citep{lifschitz_local_1991}:
\begin{equation}
\label{eq:WKB_amplitude_1}
\left\lbrace
\begin{array}{rl}
\boldsymbol{\mathcal{K}} ~&=~ \boldsymbol{\nabla} \phi \\
\displaystyle \frac{\mathrm{d} \boldsymbol{\mathcal{K}} }{\mathrm{d} t}~ &= - ~  \mathsfbi{B}^T~ \boldsymbol{\mathcal{K}} \\
\displaystyle \frac{\mathrm{d} \boldsymbol{a} }{\mathrm{d} t} 
~&=~ \displaystyle \left(2 \frac{ \boldsymbol{\mathcal{K}} \boldsymbol{\mathcal{K}}^T}{\boldsymbol{\mathcal{K}}^2} - \mathsfbi{I} \right) \mathsfbi{B} \boldsymbol{a} 
- 2 \left( \frac{ \boldsymbol{\mathcal{K}} \boldsymbol{\mathcal{K}}^T}{\boldsymbol{\mathcal{K}}^2} - \mathsfbi{I}  \right) \left(\boldsymbol{e}_z \times \boldsymbol{a} \right)
- N^2 \left( \frac{ \boldsymbol{\mathcal{K}} \boldsymbol{\mathcal{K}}^T}{\boldsymbol{\mathcal{K}}^2} - \mathsfbi{I}  \right) \Theta \boldsymbol{e}_s \\
\displaystyle \frac{\mathrm{d} \Theta }{\mathrm{d} t} ~&=~ - \tilde{\boldsymbol{S}} (t) \cdot \boldsymbol{a}  
\end{array}
\right. .
\end{equation}
The equation on $\boldsymbol{\mathcal{K}}$ can be solved analytically: it follows the same time evolution as $\tilde{\boldsymbol{S}}(t)$, $\boldsymbol{\mathcal{K}} = \hat{\mathsfbi{R}} (t) \boldsymbol{\mathcal{K}}_0$ where $\hat{\mathsfbi{R}} (t) $ has been defined in the time evolution of stratification (\ref{eq:stratification_time}) and $\boldsymbol{\mathcal{K}}_0$ is an initial condition vector.

\begin{figure}
\centering
\includegraphics[width=0.5\linewidth]{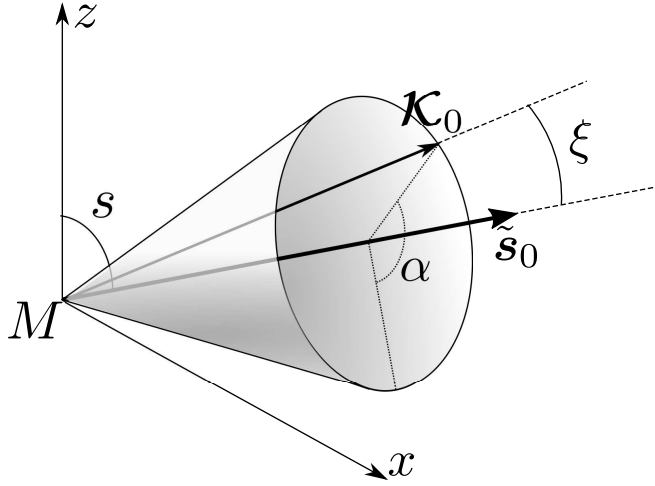}
\caption{Schematic diagram defining the relevant angles to describe the relative position between initial stratification $\tilde{\boldsymbol{s}}_0$ and wave-vector $\boldsymbol{\mathcal{K}}_0$. The angle between $\tilde{\boldsymbol{s}}_0$ and the axis $z$ is $s$ defined in figure \ref{fig:stratif_time}.}
\label{fig:def_angle_stratif}
\end{figure}

\begin{figure}
\centering
\includegraphics[width=0.45\linewidth]{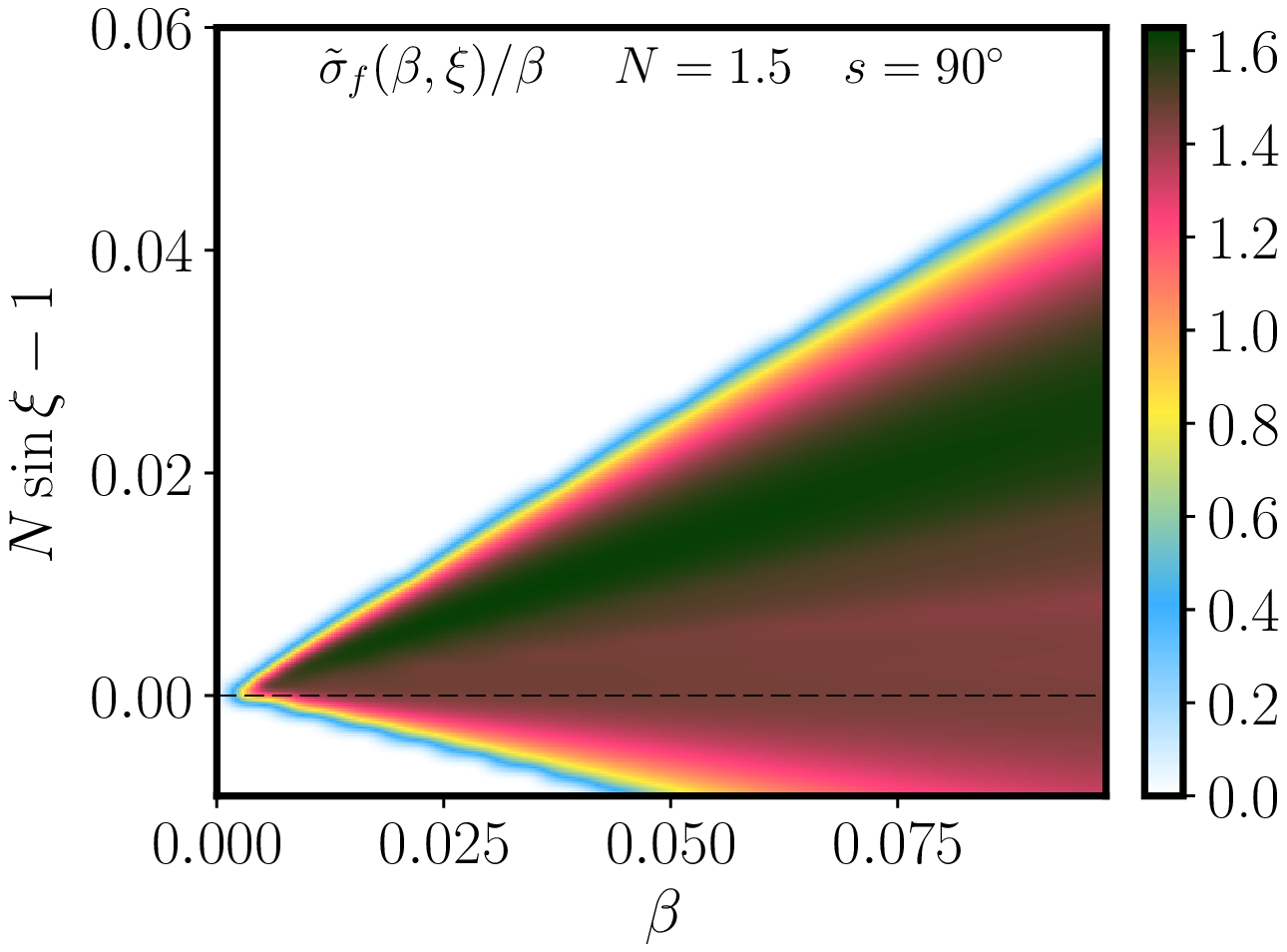}
\includegraphics[width=0.45\linewidth]{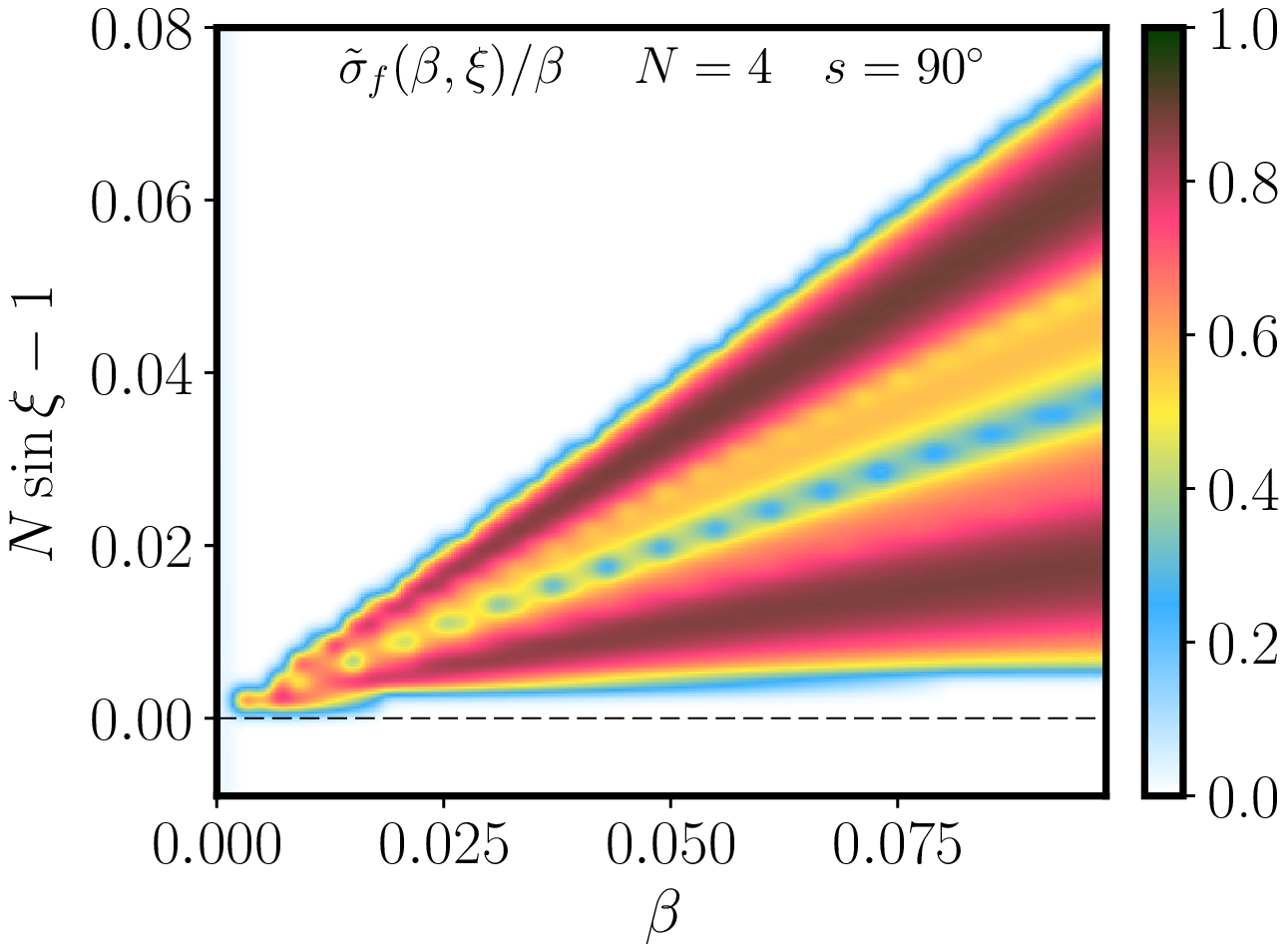}
\caption{Map of the growth rate computed via the Floquet theory $\tilde{\sigma}_f$ as a function of the ellipticity $\beta$ and the angle $\xi$ defined in figure \ref{fig:def_angle_stratif} and equation (\ref{eq:def_angles}) for \textbf{left:} $N= 1.5$ and $s= 90^\circ$ and \textbf{right:} $N= 4$ and $s= 90^\circ$. As it will be shown hereafter, the growth rate is linearly growing with the ellipticity. We therefore normalise the growth rate by $\beta$ which allows to identify the limits of the Floquet resonance tongue. The latter converges towards $N \sin \xi = \gamma = 1$ for $\beta \rightarrow 0$ showing that the resonance is subharmonic. Note that the higher growth rate area is always above the dashed line $N \sin \xi = 1$. } 
\label{fig:floquet_tongue}
\end{figure}

At the lowest order ($\beta = 0$), the linear operators $\mathsfbi{B}$ and $\boldsymbol{e}_z \times \cdot $ are equal. 
Since the shearing and stretching effects are entirely due to the ellipticity of streamlines, the stratification $\tilde{\boldsymbol{S}}$ and wave-vector $\boldsymbol{\mathcal{K}}$ have a purely rotating motion at rate $\gamma$.  
Taking the time derivative of the last equation in (\ref{eq:WKB_amplitude_1}) gives a second order differential equation:
\begin{equation}
\label{eq:WKB_beta0_1}
\displaystyle \frac{\mathrm{d}^2 \Theta }{\mathrm{d} t^2} 
- N^2 \left( \frac{(\tilde{\boldsymbol{s}}_0 \cdot \boldsymbol{\mathcal{K}}_0)^2 }{\boldsymbol{\mathcal{K}}_0^2} -1 \right) \Theta = 0 
\end{equation}
where $\tilde{\boldsymbol{s}}_0$ is the unit vector defining the initial stratification direction, as defined in (\ref{eq:stratif_adim}).
It is then convenient to introduce $\xi$, the angle between the initial stratification and wave-vector as represented in figure \ref{fig:def_angle_stratif}.
$\Theta$, and consequently the velocity component sensitive to the stratification $\boldsymbol{a} \cdot \tilde{\boldsymbol{S}}$, oscillate at a frequency $\lambda = \pm N \sin \xi$. 
At the lowest order, the internal wave dispersion relation is retrieved.
To further analyse the linear growth of the instability and to obtain quantitative growth rates, we perform the Floquet analysis of the system (\ref{eq:WKB_amplitude_1}) \citep{bender_advanced_1978}. 
This can be done since the vectors $\boldsymbol{\mathcal{K}}$, $\tilde{\boldsymbol{S}}$ and $\boldsymbol{e}_s$ all oscillate at the same frequency $\varpi$, defined in (\ref{eq:def_omega_E}). 
The linear operator RHS of (\ref{eq:WKB_amplitude_1}) therefore oscillates with a period $2 \pi / \varpi$. 
Knowing the time evolution of $\boldsymbol{\mathcal{K}}$, we solve the differential equations over the vector $( \boldsymbol{a}, \theta)$ from $t=0$ to $t = T= 2\pi / \varpi$. 
The initial condition is the identity matrix $\mathsfbi{I}$. 
The final value $( \boldsymbol{a}(T), \theta(T))$ for each initial condition corresponds to a monodromy matrix $\mathsfbi{\Phi}$ and the growth rate $\sigma$ of the instability is then related to its eigenvalues $\mu_i$ such that \citep{bender_advanced_1978,cebron_libration-driven_2014}:
\begin{equation}
\label{eq:WKB_res_mu}
\sigma = \frac{\varpi}{2 \pi}  \max \left\lbrace \ln \mu_i \right\rbrace ~~~ \mbox{or} ~~~ \tilde{\sigma} = \varpi \max \left\lbrace \ln \mu_i \right\rbrace
\end{equation}
where $\tilde{~\cdot~}$ refers to a growth computed per tidal cycles. 
As the resonant wave-vector is \textit{a priori} unknown, the Floquet analysis is performed for different $\boldsymbol{\mathcal{K}}_0$. 
This initial wave vector is parameterised by its norm $\mathcal{K}_0$ (which does not play any role in the inviscid limit), the angles $s$, $\xi$ and $\alpha$ represented in figure \ref{fig:def_angle_stratif}, such that:
\begin{equation}
\label{eq:def_angles}
\boldsymbol{\mathcal{K}}_0 = \mathcal{K}_0
\left[
\begin{array}{c}
 \sin \xi ~ \cos \alpha  ~ \cos s ~ + ~ \sin s ~\cos \xi  \\
  \sin \xi ~ \sin \alpha  \\
  -\sin \xi ~ \cos \alpha  ~ \sin s ~ + ~ \cos s ~\cos \xi  \\
\end{array}  \right]  ~.
\end{equation}
Such a parameterisation merely comes from the expression of $\boldsymbol{\mathcal{K}}_0$ in the spherical coordinates $(\alpha, \xi)$ with a polar axis $(M, \tilde{\boldsymbol{s}}_0 )$ (see figure \ref{fig:def_angle_stratif}). 
With $s$ used as a control parameter, resonance is found exploring the values of $\tilde{\sigma}$ in the $(\xi,\alpha)$ space. 
The range of angles $\xi$ to explore is non-trivial.
We show in figure \ref{fig:floquet_tongue} the maximum (respectively to $\alpha$) growth rate computed with the Floquet theory as a function of $\xi$ and $\beta$.
The non-zero growth rate area delimits Floquet tongues which stretch towards $N \sin \xi = \gamma = 1$ as $\beta$ goes to $0$. 
The resonant waves are therefore parametrically excited close to half the frequency of the forcing flow.
Note however that the Floquet tongues are not symmetric around $N \sin \xi = 1$: the areas with maximum growth rates are always slightly above this line. 
As a consequence, to compute the theoretical maximum growth rate, we explore a range of angle $\xi$ around $ \arcsin(1/N)$ with a tolerance of order $\beta$.
Note that the theoretical growth rate could also have been analytically computed via a multi-scale analysis where $\beta t$ is the slow time. 
This would have given the asymptotic resonant values of $\alpha$ for $\beta \rightarrow 0$. 
However, the complexity of the first order operator respective to $\beta$ is such that the problem might be intractable. 
This might be due to the low degree of symmetry as the angle between the orbital plane and the stratification axis is arbitrary.

\subsection{Comparison between DNS and linear stability analysis}

\begin{figure}
\includegraphics[width=\linewidth]{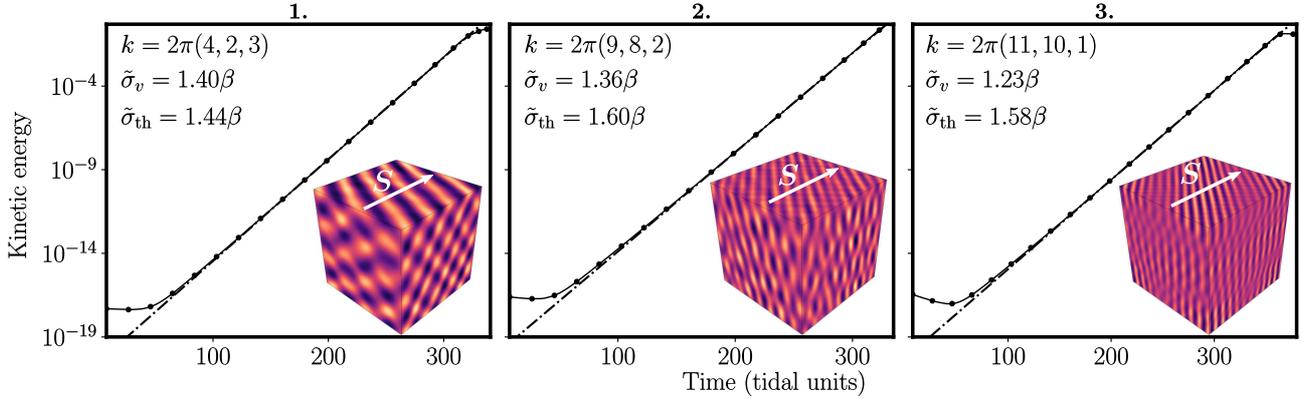}
\caption{Growth of the instability for $\beta = 5 \times 10^{-2}$, $N= 1.5$ and $s = 90^\circ$ initiated from broad-band white noise with $k / (2 \pi)$ in three different intervals \textbf{1:}$\left[ 5, 10 \right]$, \textbf{2:}$\left[ 10, 14 \right]$ and \textbf{3:}$\left[ 14, 20 \right]$. 
The viscous growth rate is obtained by a linear fitting in log-lin coordinates (shown as dashed-dotted lines). 
The corresponding snapshot of the buoyancy perturbation field is given for each experiment with an indication of the stratification direction. 
It is used to determine the growing mode.}
\label{fig:DNS_growth}
\end{figure}

We perform direct numerical simulations setting $\beta= 0.05$, $N = 1.5$, $s = 90^\circ$ with a Reynolds number $Re = 10^{6.5}$.  
A first try was initiated from a broadband white noise with $k/(2 \pi) \in \left[ 4, 20 \right]$.
The kinetic energy displays an exponential growth but snapshots reveal several entangled growing modes. 
To better quantify the growth rate and modes selection, we then restrict the broad-band noise to three intervals \textbf{1:}$\left[ 5, 10 \right]$, \textbf{2:}$\left[ 10, 14 \right]$ and \textbf{3:}$\left[ 14, 20 \right]$ which allows to isolate growing modes with approximately the same wavelength.  
The kinetic energy corresponding to these three initial conditions is shown in figure \ref{fig:DNS_growth} with snapshots of the buoyancy perturbation field $\vartheta$ during the exponential growth phase. 
In each case, the kinetic energy is exponentially growing and the buoyancy field bears a plane wave structure, confirming that the instability mechanism is based on wave resonance.
These DNS results allow to calculate the viscous growth rate $\tilde{\sigma}_v$ in tidal units. 
The inviscid growth rate $\tilde{\sigma}_{\rm{th}}$ (expressed in tidal cycles) is then obtained by substracting the viscous damping of the growing mode, \textit{i.e.} $\tilde{\sigma}_{\rm{th}} = \tilde{\sigma}_{v} + 2 \pi k^2 Re^{-1} $ with $k$ the wave number of the mode (as $\tilde{\sigma}_{\rm{th}}$ is in tidal units, a $2 \pi$ factor must be added to the damping rate). 
\begin{figure}
\centering
\includegraphics[width=0.8\linewidth]{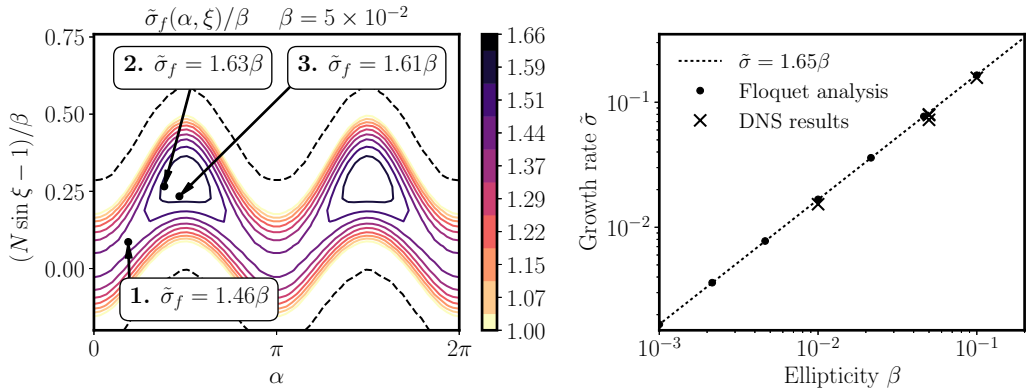}
\caption{\textbf{Left:} map of the growth rate $\tilde{\sigma}_f$ computed with the Floquet analysis as a function of the angles $\xi$ and $\alpha$. 
The black dots correspond to the location of the growing modes observed in figure \ref{fig:DNS_growth} for which the theoretical $\tilde{\sigma}_f$ growth rate is given. 
The dashed black lines highlight the marginal stability.  
\textbf{Right:} comparison between systematic calculation of the maximum growth rate from the Floquet analysis and DNS results.
The growth rates $\tilde{\sigma}_f$ computed with Floquet theory are all aligned on $\tilde{\sigma}_f  = 1.65 \beta$. }
\label{fig:synthese_benchmark}
\end{figure}
These results are then compared to the theoretical inviscid growth rate $\tilde{\sigma}_f$ given by the Floquet analysis of equations (\ref{eq:WKB_amplitude_1}). 
The map of $\sigma_f(\alpha,\xi)$ is displayed in figure \ref{fig:synthese_benchmark} where the location of the growing mode for each DNS is highlighted by a black dot associated to the corresponding $\tilde{\sigma}_f$. 
We first note that, both in DNS and theory, the angle $\xi$ satisfies the condition $N \sin \xi = 1$ with a tolerance smaller than $\beta$, as it was expected from qualitative arguments developed in the preceding paragraph. 
In addition, the theoretical growth rate $\tilde{\sigma}_f$ is close to the growth rate $\tilde{\sigma}_{\rm{th}}$ measured in DNS with a relative difference less than $2$\% . 
With this very good agreement between DNS and the linear theory, we can now analyse the dependence of the growth rate on the control parameters using rapid linear theory only.
\subsection{Linearity with the ellipticity $\beta$}
The amplitude of the periodic stretching and shearing responsible for the parametric excitation of internal waves is proportional to the ellipticity $\beta$. 
Another way to validate the DNS and the linear stability analysis is to examine the consequent expected linearity in $\beta$ of the growth rate as in the case of tidally-driven elliptical instability \citep{le_dizes_three-dimensional_2000,
kerswell_elliptical_2002,
grannan_tidally_2017}.
As shown in figure \ref{fig:synthese_benchmark} (right panel), the theoretical maximal growth rate $\tilde{\sigma}_f$ inferred from Floquet theory is very well described by a linear function over several orders of magnitude, in the case with $N = 1.5$ and $s = 90^\circ$. 
The growth rate computed from DNS  is in addition very close to this theoretical line. 
\subsection{Dependence on the stratification angle $s$}
\begin{figure}
\centering
\includegraphics[width=0.8\linewidth]{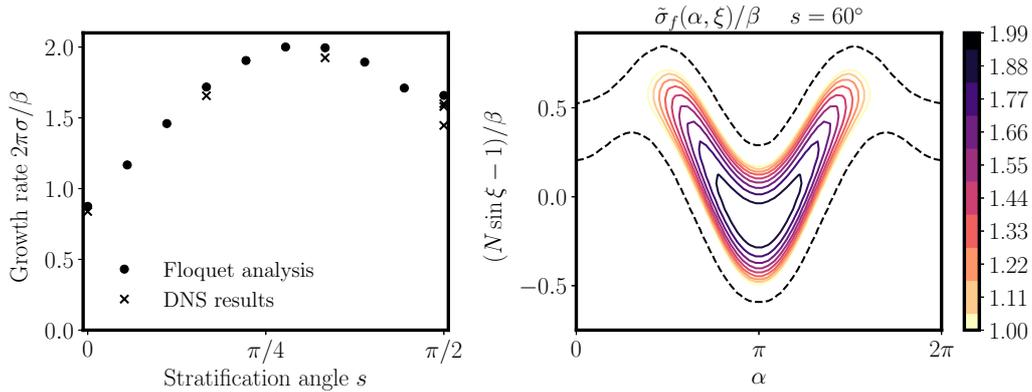}
\caption{\textbf{Left:} theoretical and numerical growth rates with $\beta= 5 \times 10^{-2}$ and $N = 1.5$ varying the stratification angle $s$. \textbf{Right:} corresponding map of the growth rate $\tilde{\sigma}_f$ computed with Floquet theory for the case $s = 60^\circ$. The black dashed lines highlight the marginal stability.}
\label{fig:s_growth_rate}
\end{figure}

To illustrate the dependence on the latitude or equivalently the stratification angle $s$ (see figures \ref{fig:stratif_time} and \ref{fig:def_angle_stratif}), we show in figure \ref{fig:s_growth_rate} the maximum theoretical growth rate of the instability for fixed $N = 1.5$. 
The main conclusion is that this instability can be triggered at any latitude in a planetary fluid layer. For this Brunt-V\"ais\"al\"a frequency, the growth is optimal between roughly $50^\circ$ and $60^\circ$. 
The mode selection of the parametric instability changes with latitude. 
Unlike the mode selected in the case $N = 1.5$ and $s= 90^\circ$ where $\alpha \simeq \pi/2$ (see figure \ref{fig:synthese_benchmark}), the mode selected at $s = 60^\circ$ lies in the plane $(xMz)$, \textit{i.e.} $\alpha \simeq \pi$ (see figure \ref{fig:def_angle_stratif}).

\subsection{Dependence on the Brunt-V\"ais\"al\"a frequency $N$ }

\begin{figure}
\centering
\includegraphics[width=0.8\linewidth]{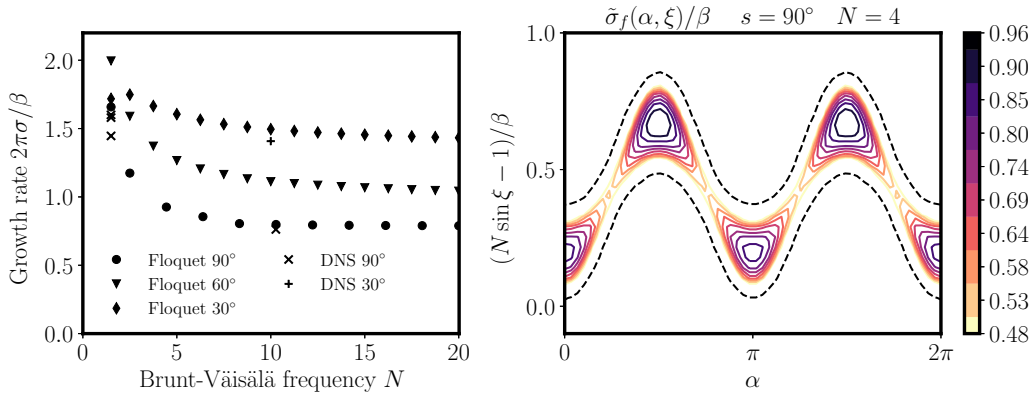}
\caption{\textbf{Left:} evolution of the  maximum growth rate for $\beta= 5 \times 10^{-2}$ at three different stratification angles; a few simulations have been carried out to confirm the agreement between theory and numerics. \textbf{Right:} map of the growth rate computed from the Floquet analysis in the case $N = 4$ with $s = 90^\circ$. The black dashed lines highlight the marginal stability }
\label{fig:N_growth_rate}
\end{figure}

The growth rate of the instability is also a function of the Brunt-V\"ais\"al\"a frequency. 
As shown in figure \ref{fig:N_growth_rate}, it tends to a limit value when $N \gg 1$.
The consequence is that the instability can be triggered at any Brunt-V\"ais\"al\"a frequency provided that it is larger than $1$ in tidal units. 
Note that for large $N$, the selected modes' wave vectors draw closer to the stratification direction as $\sin \xi \sim 1/N$.

\subsection{Conclusion}

The linear stability analysis examined theoretically via WKB analysis and Floquet theory is quantitatively consistent with the results of direct numerical simulations. 
This first study has two consequences. 
It allows to assert that provided the dissipation is low enough, a parametric excitation of internal waves can be excited in a planetary fluid layer undergoing homogeneous tidal deformations. 
This instability can be triggered at any latitude. 
However, the mode selection seems difficult to predict as it depends on latitude. 
At least it is confirmed that the growing waves are selected because their frequencies are close to half the forcing frequency $2 \gamma$. 
In addition, this first linear study validates the use of local direct numerical simulations under a shearing box approximation with time-varying wave-vectors, as this method is in excellent agreement with the linear WKB-Floquet theory. 
\section{Non-linear saturation of the instability}

To further analyse this tidally-driven instability of internal waves, we now focus on its non-linear saturation. 
This regime is the most relevant to the understanding of natural systems, in particular to comprehend the dissipation rate of the input tidal energy and the turbulent mixing in the oceans, or to study the possible existence of dynamo action in stably stratified planetary cores. 
Although we cannot address here all those issues, we strive to exhibit the key features of the non-linear saturation of this tidally-driven instability as a basis for future works. 
We performed many simulations, all with an ellipticity $\beta = 5 \times 10^{-2}$, to explore the influence of the Reynolds number $Re$, the Brunt-V\"ais\"al\"a frequency $N$, and the latitude or stratification angle $s$ in the low forcing intensity and low dissipation regime relevant to geophysics. 
The Prandtl (or Schmidt) number $Pr$ is kept constant at $Pr = 1$. 
These simulations are all summed up in table \ref{tab:runs_statistical} where the input parameters are referenced along with the main output statistical quantities.

\subsection{Sustained instability and turbulence}
\begin{table}
  \begin{center}
\def~{\hphantom{0}}
  \begin{tabular}{cccc|cccccc}
   \multicolumn{4}{c|}{Input variables} 
   &  \multicolumn{6}{c}{Output variables} \\[3pt] 
    Resolution & $N$  & $ \log(Re) $   &   $ s $ & $k_{\mathrm{res}}/(2\pi)$ & $ u_{\mathrm{rms}}~ (\times 10^{-3}) $ & $Fr$ & $\varepsilon_{\rm{k}} ~ (\times 10^{-8})$ & $Re_o$ & $\mathcal{R}$ \\[3pt]
      $256^3$ & 1.5   & 6.0 & $45^\circ$ & 12.0 & $4 \pm 1 $ & $0.032$ & $5 \pm 3 $ & $ 332$ & $0.35$ \\
      $256^3$ & 1.5   & 6.5 & $45^\circ$ & 12.0 & $4.8 \pm 0.4 $ & $0.039$ & $3 \pm 1 $ & $ 1279$ & 2.0 \\
      $256^3$ & 1.5   & 6.75 & $45^\circ$ & 12.0 & $4.4 \pm 0.4 $ & $0.035$ & $2.1 \pm 0.7 $ & $ 2049$ & 2.6 \\
      $512^3$ & 1.5   & 7.0 & $45^\circ$ & 12.0 & $4.4 \pm 0.3 $ & $0.035 $ & $1.8 \pm 0.1$ & $ 3673$ & 4.7\\
      $256^3$ & 1.5   & 6.5 & $0^\circ$ & 5.4 & $3.7 \pm 0.7 $ & $0.030$ & $1.3 \pm 0.6   $ & $ 980$ & 0.9\\
      $256^3$ & 1.5   & 6.5 & $90^\circ$ & 11.2 & $5.9 \pm 0.6 $ & $0.047 $ & $4 \pm 1    $ & $ 1550$ & 3.6\\
      $256^3$ & 2.0   & 6.5 & $90^\circ$ & 5.8 & $5.4 \pm 0.5 $ & $0.016 $ & $3 \pm 1    $ & $ 2950$ & 0.74\\
     $256^3$ & 4.0   & 6.5 & $90^\circ$ & 14.6 & $2.3 \pm 0.9 $ & $0.008 $ & $ 1  $ & $ 500$ & 0.03\\
      $256^3$ & 4.0   & 7.0 & $90^\circ$ & 14.6 & $3.3 \pm 0.5 $ & $0.012 $ & $ 1.5 \pm 0.2  $ & $ 720$ & 0.10\\
  \end{tabular}
  \caption{
Input parameters and measured statistical properties of the flow for each simulation. 
$k_{\mathrm{res}}$ is the principal wave number of the resonant modes emerging during the growth phase.
$u_{\rm{rms}}$ is the rms velocity computed from the mean of the kinetic energy. 
The Froude number $Fr$ is computed as $u_{\mathrm{rms}}/(N\lambda_{\rm{res}})$.
$\varepsilon_{\rm{k}}$ is the saturation dissipation rate defined as $-Re^{-1} \left\langle \partial_i u_j \partial_i u_j \right\rangle$ summed over the whole box.   
The output Reynolds $Re_o$ number and the buoyancy Reynolds number $\mathcal{R}$ are respectively defined in equations (\ref{eq:output_reynolds}) and (\ref{eq:buoyancy_Reynolds}).
When errorbars are given, they correspond to the variance of the quantity over the total duration of the saturation phase. 
Note that $N = 4$ and $Re = 10^{6.5}$ is intermittently turbulent and the Reynolds number had to be pushed up to $10^7$ to observe a sustained turbulence.}
  \label{tab:runs_statistical}
  \end{center}
\end{table}

\begin{figure}
\centering
\includegraphics[width= 0.45\linewidth]{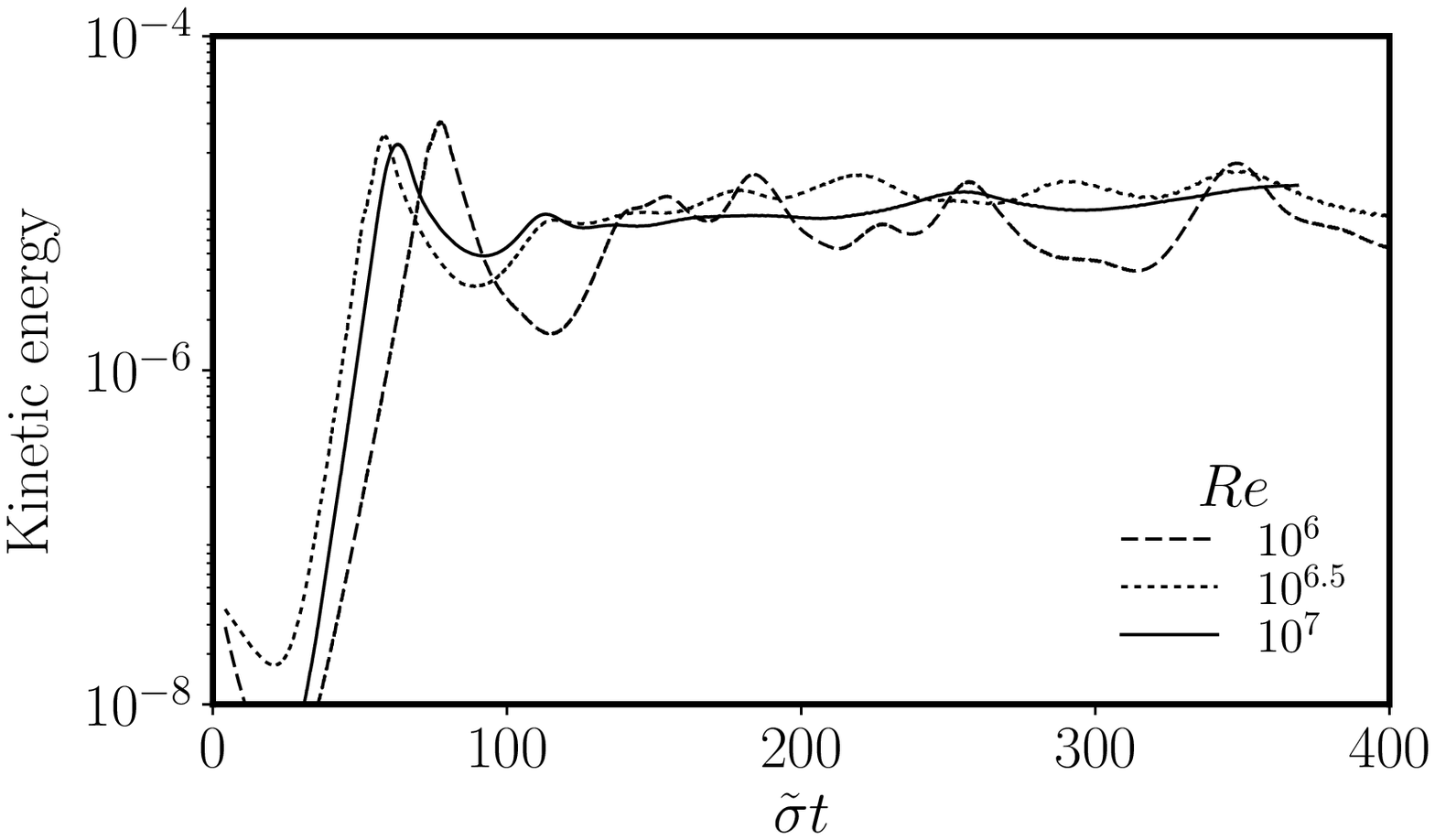}
\includegraphics[width= 0.45\linewidth]{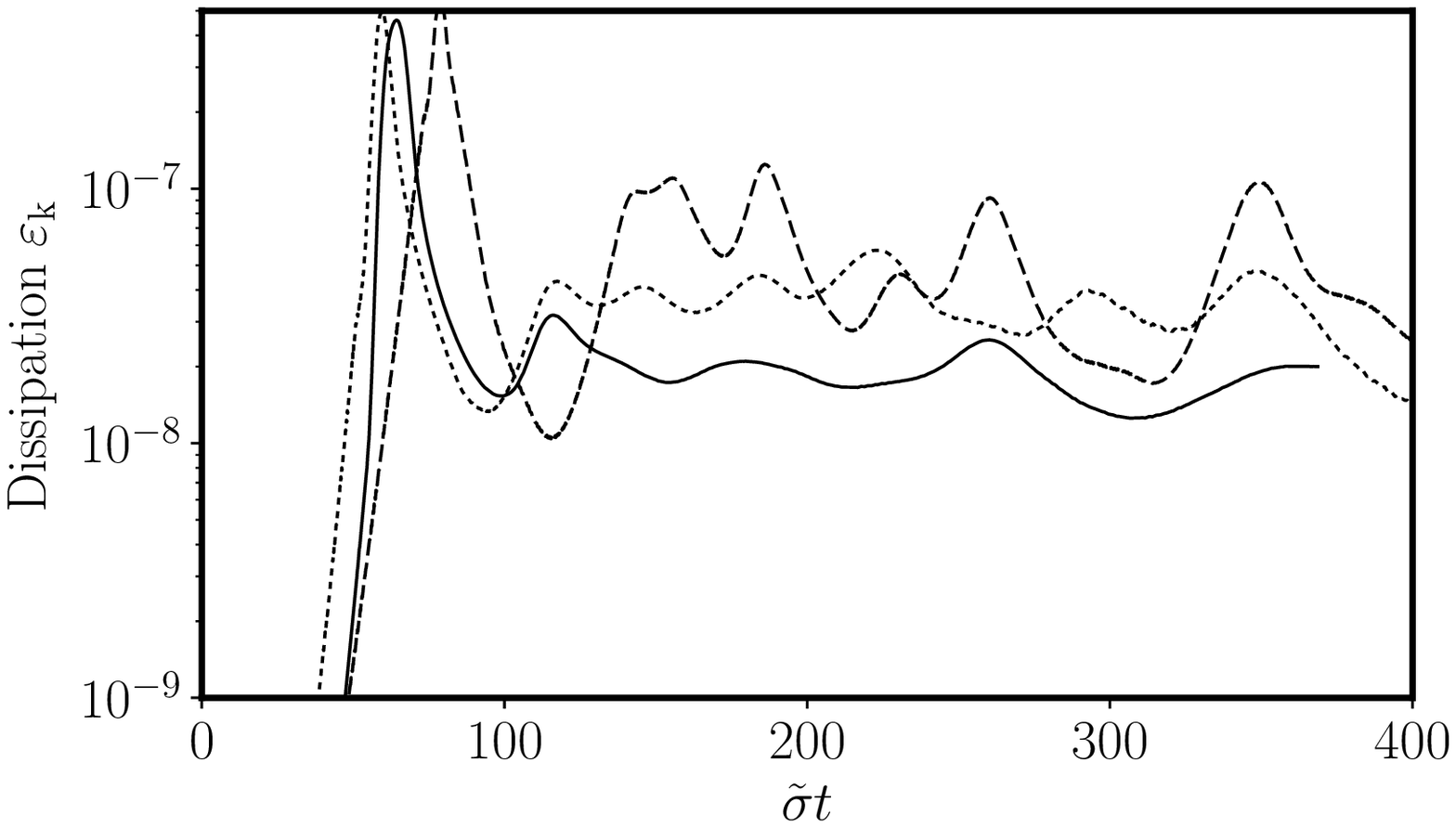}
\includegraphics[width= 0.90\linewidth]{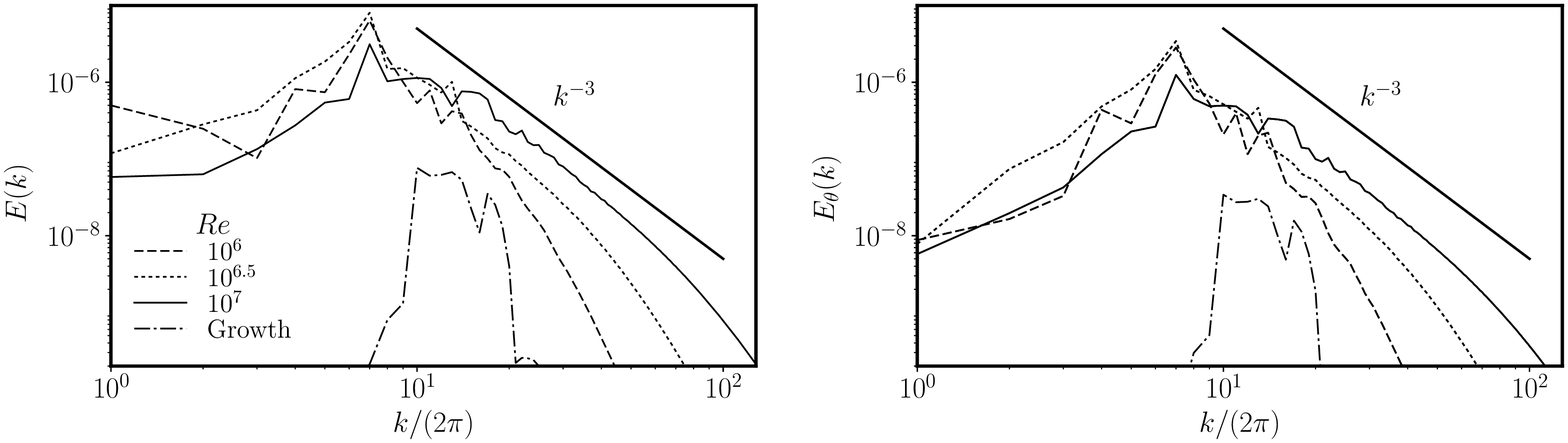}
\caption{\textbf{Top:} Time evolution of the kinetic energy and the dissipation $\varepsilon_{\rm{k}}$ when $Re$ is varied while keeping constant $s= 45^\circ$ and $N = 1.5$. 
The time is normalised by the theoretical growth rate $\sigma = \tilde{\sigma}_f$. 
\textbf{Bottom:} corresponding power spectra of the velocity $E(k)$ and of the buoyancy $E_\theta (k)$; they are averaged for $\tilde{\sigma} t \in \left[150, 400 \right]$. 
The black solid line materialises a $k^{-3}$ power law.  
The dashed-dotted spectra is computed during the growth phase to show that the energy is primarily injected in a narrow band  of wave numbers through the instability mechanism. }
\label{fig:system_Re_varied}
\end{figure}
\begin{figure}
\centering
\includegraphics[width=0.90\linewidth]{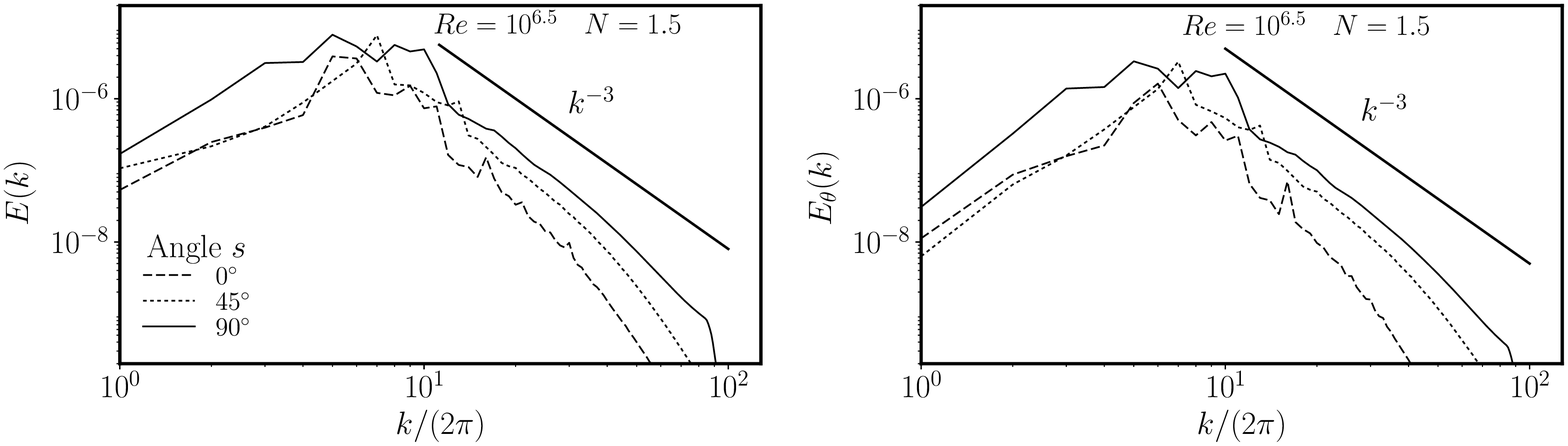}
\includegraphics[width=0.90\linewidth]{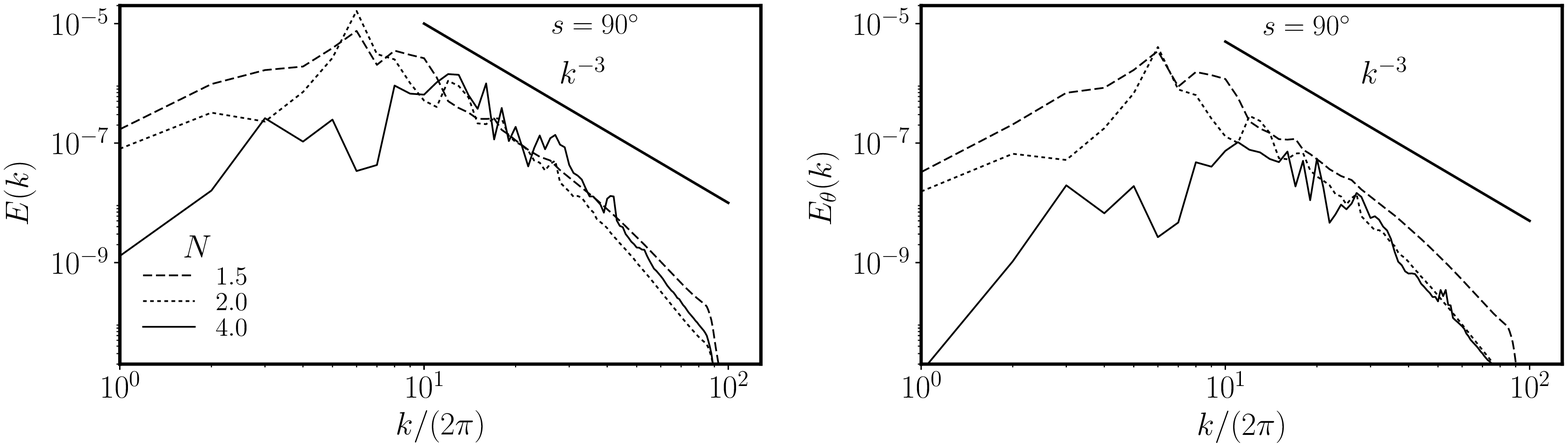}
\caption{ \textbf{Top:} power spectra of the velocity $E(k)$ and of the buoyancy $E_\theta (k)$ for $s \in \left\lbrace 0^\circ, 45^\circ, 90^\circ\right\rbrace$ at $Re = 10^{6.5}$ and $N = 1.5$. \textbf{Bottom:} the same quantities for $s=90^\circ$ with $N \in \left\lbrace 1.5, 2, 4\right\rbrace$. The Reynolds number is $10^{6.5}$ except for $N=4$ where it had to be increased to $ 10^7$ to observe sustained turbulence.  }
\label{fig:spectra_N_s_varied}
\end{figure}
\begin{figure}
\centering
\includegraphics[width=0.35\linewidth]{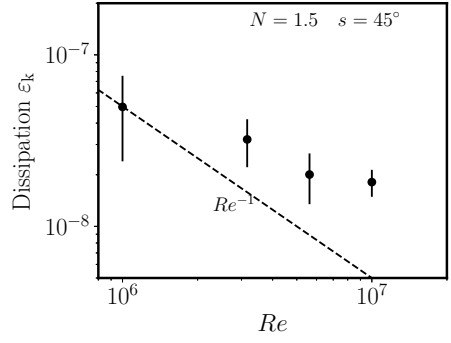}
\caption{Mean value of the dissipation rate $\varepsilon_{\rm{k}}$  as a function of the Reynolds number for $N = 1.5$ and $s= 45^\circ$. The errorbars account for the standard deviation of $\varepsilon_{\rm{k}}$ during the saturation phase.  }
\label{fig:dissipation_Re}
\end{figure}

\begin{figure}
\centering
\includegraphics[width=0.85\linewidth]{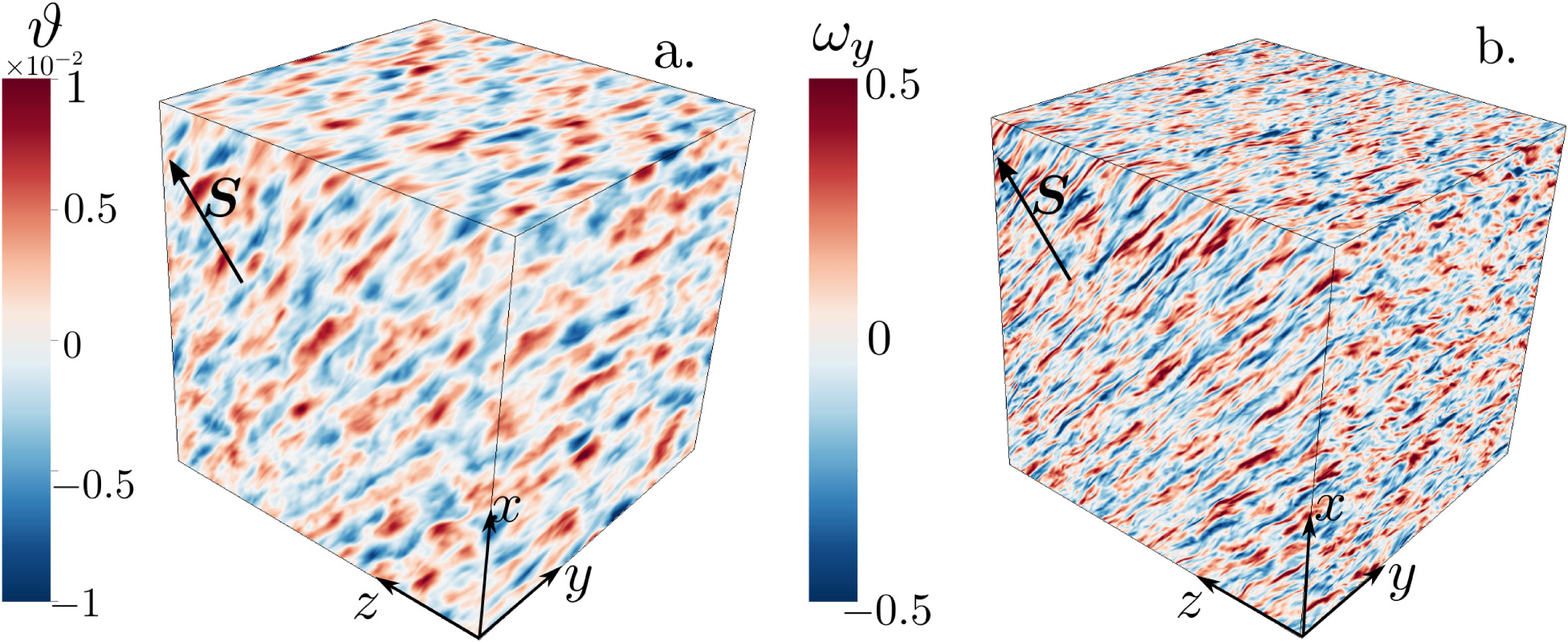}
\includegraphics[width=0.85\linewidth]{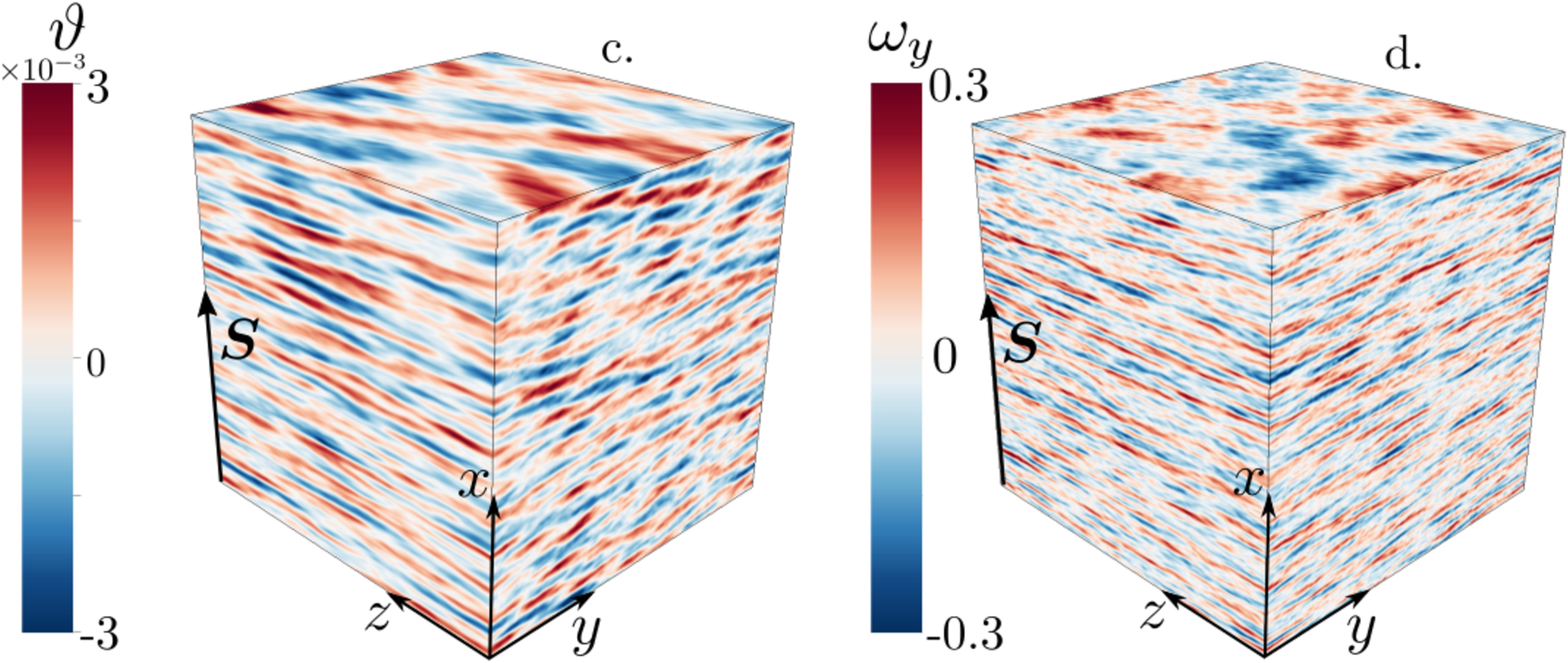}
\caption{Typical snapshots of the buoyancy field (\textbf{a.} and \textbf{c.}) and the $y$ component of the vorticity (\textbf{b.} and \textbf{d.}) in the saturated phase for \textbf{a.}, \textbf{b.}: $N = 1.5$, $s = 45^\circ$ and $Re = 10^7$; \textbf{c.}, \textbf{d.}: $N = 4$, $s = 90^\circ$ and $Re = 10^7$. }
\label{fig:snapshot_buy_vorticity}
\end{figure}

As in the linear stability analysis, the non-linear fate of the instability is mainly tracked via the time evolution of the total kinetic energy in the Lagrangian box. 
Figure \ref{fig:system_Re_varied} shows that once the instability has reached saturation, the kinetic energy is maintained throughout time for the considered parameters. 
As indicated by figure \ref{fig:system_Re_varied}, changing the Reynolds number from $10^{6}$ to $10^{7}$ does not influence the time-averaged value of the kinetic energy provided that $N$ and $s$ are kept constant. 
We note though that the variations around the mean energy level are larger for the lowest values of the Reynolds number. 
This is reminiscent of cyclic resonance and turbulence breakdown often occurring in systems close to the instability threshold. 
It has been observed for instance for the elliptical instability in rotating fluids \citep{grannan_experimental_2014,favier_generation_2015} or in the case of parametrically excited internal waves in a Faraday instability setup \citep{benielli_excitation_1998}. 
To examine whether the saturation flow is influenced by stratification, we compute the Froude number based on the resonant wavelength and the rms velocity. 
The resonant wavelength $\lambda_{\mathrm{res}}= 2 \pi / k_{\rm{res}}$ is an output parameter resulting from the mode selection during the growth phase (see table \ref{tab:runs_statistical}), and the rms value of the velocity in the saturated phase is computed from the saturation time-average of the kinetic energy $\overline{E_{\mathrm{k}}}$ such that $u_{\mathrm{rms}} = \sqrt{2 \overline{E_{\mathrm{k}}}}$. 
The Froude number $Fr$ is therefore defined as \begin{equation}
\label{eq:Froude}
Fr = \frac{u_{\mathrm{rms}}}{\lambda_{\mathrm{res}} N} .
\end{equation}
The values of $u_{\mathrm{rms}}$ and $Fr$ are all referenced in table \ref{tab:runs_statistical}.
As $Fr = \mathcal{O}(10^{-2})$ for all the simulated configurations, we conclude that the background stable stratification strongly affects the saturated flow. 

In addition, the saturation of the instability is associated with the spontaneous creation of small scales.
The isotropic power spectrum of both velocity ($E(k)$) and buoyancy ($E_{\theta} (k)$) are shown in figures \ref{fig:system_Re_varied} and \ref{fig:spectra_N_s_varied}. 
They are computed as:
\begin{equation}
\label{eq:def_power_spectra}
E(k) = \sum_{k \leq\vert \boldsymbol{k} \vert < k+1} \vert \hat{\boldsymbol{u}}_{\boldsymbol{k}} \vert^2
~~~~ \mbox{and} ~~~
E_{\theta}(k) = \sum_{k \leq\vert \boldsymbol{k} \vert < k+1} \vert \hat{\vartheta}_{\boldsymbol{k}} \vert^2 ~ . 
\end{equation}
In the high Reynolds number limit, they converge towards a $k^{-3}$ power spectrum, independently of $N$ and $s$. 
Note that a similar velocity power spectrum has been observed by \cite{brouzet_internal_2016_thesis} in the close context of turbulence driven by a forced internal wave attractor. 
It has also been measured at low Froude number by \cite{rorai_stably_2015} in a stratified turbulence randomly forced at large scale. 
From the excitation of a few unstable internal waves, this instability mechanism manages to create sustained stratified turbulence and smaller scales. 
In addition, the apparent equipartition between velocity and buoyancy power spectra points towards a dynamics dominated by internal waves. 
To better characterise the turbulent flow resulting from the saturation of the instability, we introduce two dimensionless parameters. 
We compute an output Reynolds number $Re_o$ based on the rms velocity and the resonant wavelength such that:
\begin{equation}
\label{eq:output_reynolds}
Re_o = Re~ u_{\mathrm{rms}} ~\lambda_{\mathrm{res}}.
\end{equation}
With this output Reynolds number, we can also compute the buoyancy Reynolds number $\mathcal{R}$ defined as \citep{brethouwer_scaling_2007}:
\begin{equation}
\label{eq:buoyancy_Reynolds}
\mathcal{R} = Re_o ~  Fr^2. 
\end{equation}
It compares a scale $\ell_b$ beyond which buoyancy effects are negligible and a scale $\ell_v$ beyond which viscosity affects the flow \citep{godoy-diana_vertical_2004,brethouwer_scaling_2007}.
For instance, in the context of classical stratified turbulence, it compares the so-called Ozmidov and the Kolmogorov length scales \citep{brethouwer_scaling_2007}.
The output Reynolds number is a $\mathcal{O}(100-1000)$ but because the Froude number is small, the buoyancy Reynolds number is at most $\mathcal{O}(1)$ (see table \ref{tab:runs_statistical}). 
This means that although the flow is turbulent, there is no significant range of scales where inertia dominates over buoyancy: all the non-dissipative scales are affected by the background stratification.
This is drastically different from recent studies on forced stratified turbulence, which are mostly focused on the $\mathcal{R} \gg 1$ regime (see for instance \cite{brethouwer_scaling_2007,
bartello_sensitivity_2013,maffioli_mixing_2016,
maffioli_dynamics_2016}).
Lastly, the instantaneous dissipation rate associated with this type of turbulence is shown in figure \ref{fig:system_Re_varied} and its mean saturation value is given in table \ref{tab:runs_statistical}. They are computed in our dimensionless units as $\varepsilon_\mathrm{k} = - Re^{-1} \left\langle \partial_i  u_j \partial_i u_j \right\rangle ~ > 0$ where $\left\langle ~\cdot~ \right\rangle $ is a volume averaging operator.
Figure \ref{fig:dissipation_Re} sums up the evolution of this dissipation with the Reynolds number at constant $N$ and $s$.
The dissipation rate $\varepsilon_\mathrm{k}$ is a decreasing function of the Reynolds number in the considered range of parameters and this decay is shallower than a $Re^{-1}$ decrease.
This is an additional signature of the development of turbulence as it indicates that the velocity gradients become steeper as the input Reynolds number is decreased. 
However, in the present range of parameters accessible with reasonable computing time, no saturation of $\varepsilon_\mathrm{k}$ at high $Re$ is reached. 
As a conclusion, at large Reynolds number, the flow resulting from the saturation of this tidally-driven instability is developing over a wide range of spatial scales from an initial resonance dominated by a most unstable wavelength. 
At a given Brunt-V\"ais\"al\"a frequency, this turbulence develops at any latitude.
Typical snapshots of this turbulent state can be found in figure \ref{fig:snapshot_buy_vorticity}.
\subsection{Internal wave turbulence}

\begin{figure}
\centering
\includegraphics[width=\linewidth]{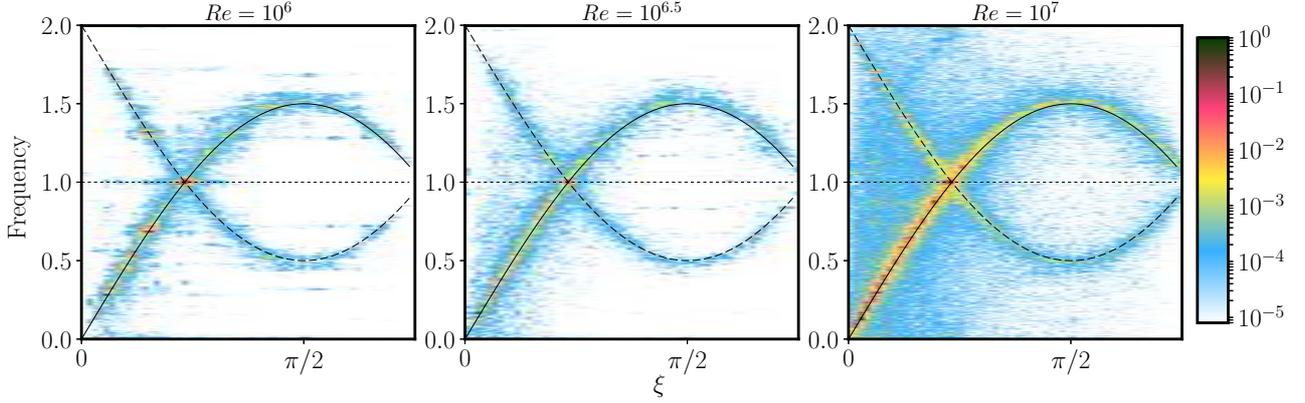}
\caption{Map of the kinetic energy as a function of the frequency of the modes and the angle $\xi$ between the stratification and the wave vector of a mode for $Re = 10^{6}$, $10^{6.5}$ and $10^{7}$ at $N=1.5$ and $s=45^\circ$ kept constant.
The Fourier transforms are performed for $\tilde{\sigma} t \in \left[200,400\right]$ and the energy is normalised by its maximum value.
The range of wave number $k$ over which the transform is processed goes from $k = 3$ to $k = 100$; changing these boundaries does not affect the map provided the most energetic scales ($k \lesssim 20$) are included.
The horizontal line represents the frequency of the first excited modes, the plain line gives the dispersion relation of the internal waves $\omega = N \sin \xi$ and the dashed line locates the modes due to non-resonant non-linear interactions between the tidal basic flow and the internal waves resulting in frequencies $2 - N \sin \xi$. 
The \textsc{Snoopy} code computes the time evolution of half the spectral space as the fields are real, the angle $\xi$ is therefore between $0$ and $\pi - \pi/4 = 3 \pi /4$.
}
\label{fig:reldisp_Revariable}
\end{figure}
\begin{figure}
\centering
\includegraphics[width=\linewidth]{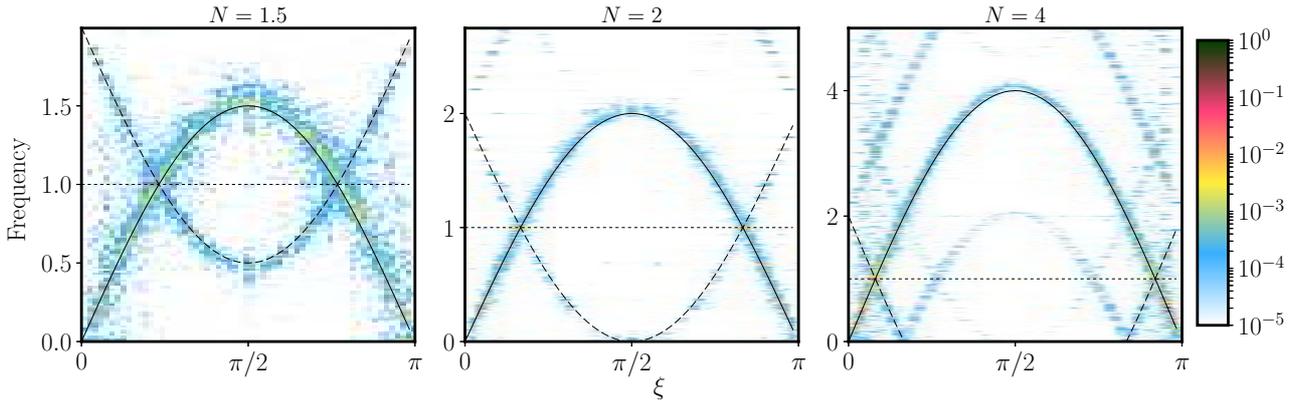}
\caption{Map of the kinetic energy as a function of the frequency of the modes and the angle $\xi$  for $N =1.5$, $2$ and $4$ with $s=90^\circ$. 
The Reynolds number is $10^{6}$ except for $N = 4$ for which it was increased to $Re = 10^{7}$ to observe sustained turbulence. 
The Fourier transforms are performed for $\tilde{\sigma} t \in \left[150,400\right]$ and the energy is normalised by the maximum value. 
Again, secondary and mirroring locations of the energy corresponding to non-resonant and nonlinear interaction of the waves and the base flow can be noticed. }
\label{fig:reldisp_Nvariable}
\end{figure}
\begin{figure}
\centering
\includegraphics[width=0.45\linewidth]{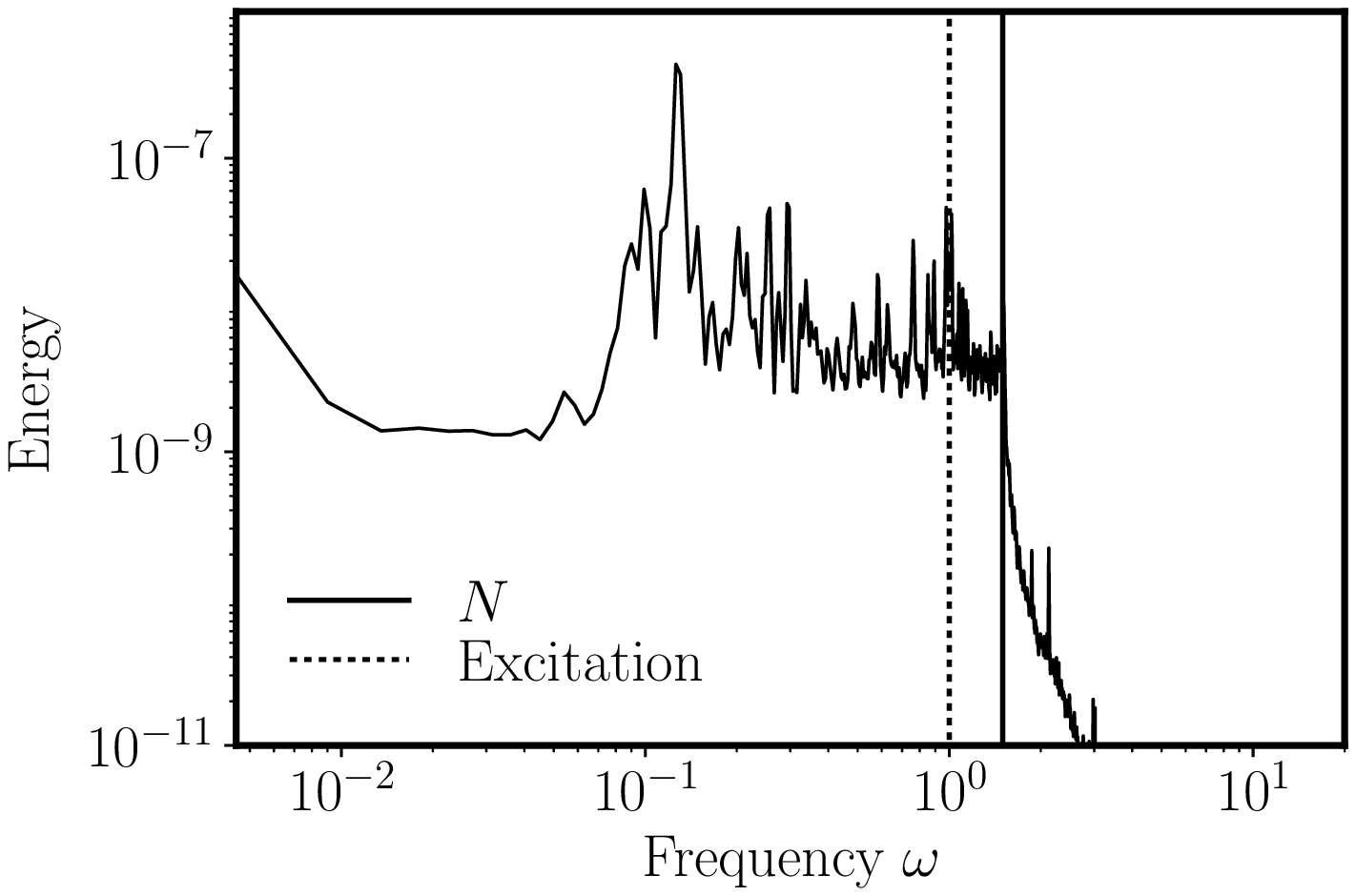}
\includegraphics[width=0.45\linewidth]{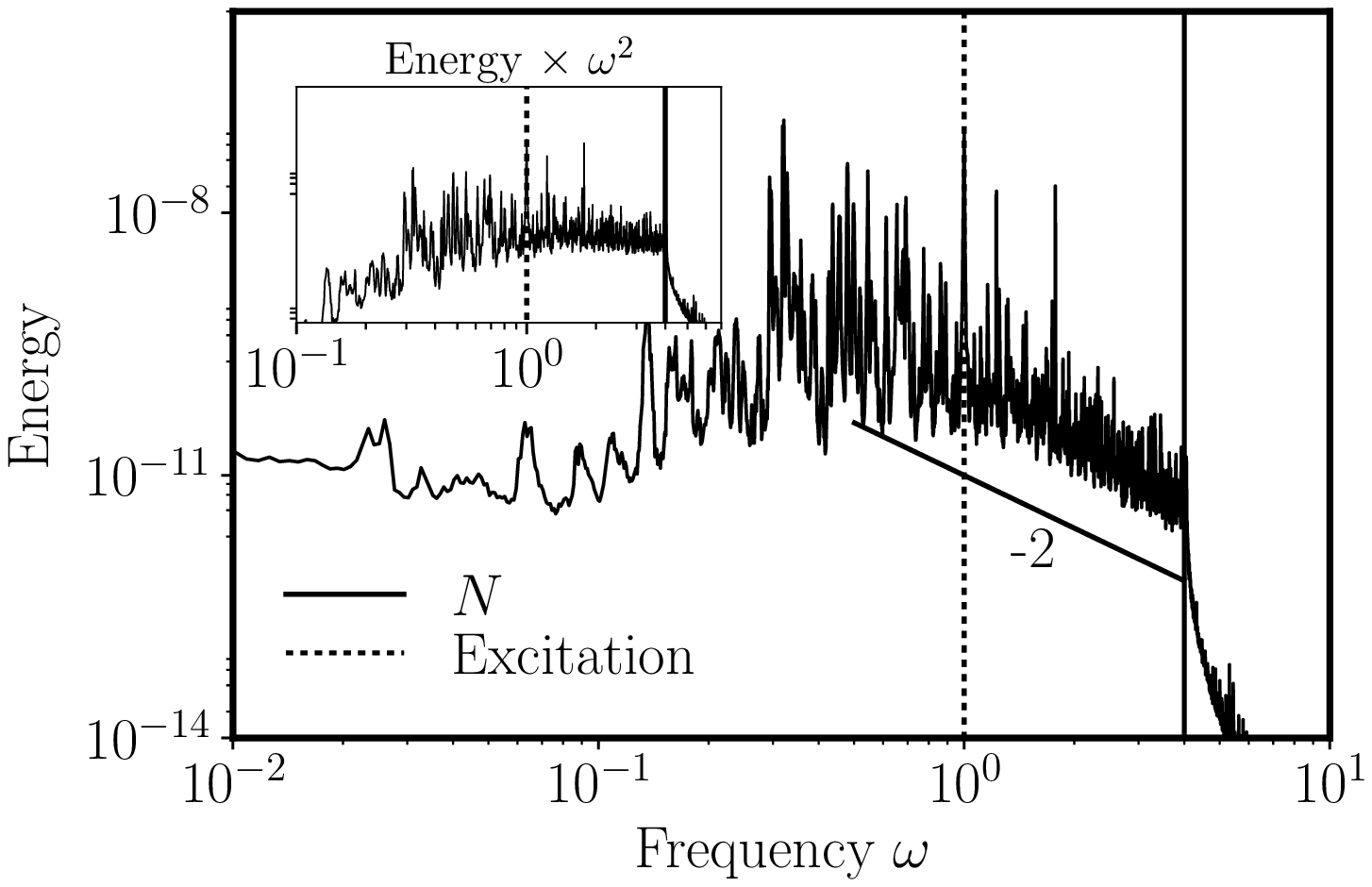}
\caption{
Temporal spectrum resulting from the acquisition of 1024 local velocity signals of all three components. 
The plain line marks the upper boundary of internal waves and the dashed line highlights the excitation frequency. 
\textbf{Left:} temporal spectrum for $N = 1.5$, $s= 45^\circ$ and $Re = 10^{7}$. 
\textbf{Right:}  temporal spectrum for $N = 4$, $s = 90^\circ$ and $Re = 10^7$. 
The insert shows the same amplitude compensated by $\omega^2$ with $\omega$ the frequency to highlight a $\omega^{-2}$ power law consistent with the high-frequency branch of the Garrett and Munk spectrum \citep{garrett_internal_1979}. 
}
\label{fig:mean_temporal_spectrum}
\end{figure}

In this paragraph, we aim at thoroughly characterising the structures generated by the non-linear saturation of the initially unstable waves. 
In simulations and experiments of stratified turbulence, the emergence of layerwise structures in which the flow is quasi-two-dimensional is frequently observed. 
These so-called ``pancakes modes" correspond to the quasi-static limit of the internal waves dynamics (\textit{i.e.} $\xi \rightarrow 0 $ and $\omega \rightarrow 0$); three-dimensional motion comes through shear instability between those layers (see \textit{e.g.} \cite{billant_self-similarity_2001,brethouwer_scaling_2007}).
Conversely, the turbulence excited by internal wave attractors leads to a different situation where the turbulence is a cascade of triadic resonances between the excited waves and a swarm of daughter waves.
It results in an internal wave turbulence \citep{brouzet_energy_2016}. 
To determine which scenario is relevant here, we propose to map the energy in the same representation as \cite{yarom_experimental_2014}, \cite{brouzet_energy_2016} and \cite{lereun_2017}, \textit{i.e.} to project the energy in the spectral space along the temporal frequency $\omega$ and the angle $\xi$ between the stratification direction $\tilde{\boldsymbol{S}}$ and the wave vector of a mode. 
This allows to determine where the energy is located around the dispersion relation of internal waves.
%
%
%
This energy map is in the present case straightforward to draw thanks to the spectral nature of our simulations. 
The flow is indeed known through the velocity in spectral space 
$\hat{\boldsymbol{u}}_{\boldsymbol{k}} (t)$. 
They can be sorted by their angles $\xi$ to obtain the quantity $\hat{\boldsymbol{u}}(\xi,t)$ defined as:
\begin{equation}
\hat{\boldsymbol{u}}(\xi_0,t) = \sum_{k,\alpha,\tilde{\xi}  \in \left[ \xi_0 , \xi_0 + \Delta \xi \right]} \hat{\boldsymbol{u}}_{\boldsymbol{k}} (t)  ~~~\mbox{with}~~~
\boldsymbol{k} = k
\left[
\begin{array}{c}
 \sin \xi ~ \cos \alpha  ~ \cos s ~ + ~ \sin s ~\cos \xi  \\
  \sin \xi ~ \sin \alpha  \\
  -\sin \xi ~ \cos \alpha  ~ \sin s ~ + ~ \cos s ~\cos \xi  \\
\end{array}  \right]  ~,
\end{equation}
where $\alpha$ is an azimutal angle, $\xi$ is a polar angle respective to the stratification axis as defined in figure \ref{fig:def_angle_stratif} and $k$ is the norm of the wave vector $\boldsymbol{k}$. 
$\Delta \xi$ is a given tolerance to assume the angle $\xi$ of a mode is equal to $\xi_0$. 
A time Fourier transform is applied to $\hat{\boldsymbol{u}}(\xi,t)$  to finally get $\hat{\boldsymbol{u}}(\xi,\omega)$.
With these definitions, we remind at this point that the dispersion relation of internal waves at a given $\boldsymbol{k}(k,\xi,\alpha)$ is $\omega^2 = N^2 \sin^2 \xi$.

The result of such a process is shown in figures \ref{fig:reldisp_Revariable} and \ref{fig:reldisp_Nvariable}. 
The most striking feature is the coincidence between the main energy locations and the dispersion relation of internal waves, similarly to \cite{yarom_experimental_2014} and \cite{lereun_2017} in rotating fluids, to \cite{brouzet_energy_2016} in stratified fluids, and to \cite{aubourg_nonlocal_2015} in the case of gravity-capillary waves. 
It confirms that, in the saturation phase, the non-linear interactions between the growing modes give rise to a cascade of daughter internal waves. 
In figure \ref{fig:reldisp_Revariable}, it can be noticed at low Reynolds number that only a few modes emerge in the non-linear saturation. 
Increasing the Reynolds number leads to filling continuously the dispersion relation.
Note that as energy is injected into the resonant modes only and as the Froude number is always small, the only way to create new waves is via a cascade of triadic resonances. 
Secondary locations of the energy mirroring the dispersion relation of internal waves can be noticed in figures \ref{fig:reldisp_Revariable} and \ref{fig:reldisp_Nvariable}. 
Their frequencies match the relation $\omega = 2 - N \sin \xi$ and are therefore associated to non-linear and non-resonant interactions between the waves of frequency $\pm N \sin \xi$ and the base flow of frequency $\pm 2$. 
The filling of the dispersion relation depends though on the Brunt-V\"ais\"al\"a frequency (figure \ref{fig:reldisp_Nvariable}). 
When $N$ is increased, modes with frequency around or below $\gamma$ seem to be more excited via nonlinear interactions than modes with frequency between $\gamma$ and $N$, at least for the Reynolds numbers  considered here (see $N=4$ in figure \ref{fig:reldisp_Nvariable}). 
In order to quantify more precisely how frequencies are excited, we propose to focus on temporal spectra obtained via the local acquisition of the three components of the velocity at several points. 
As shown in figure \ref{fig:mean_temporal_spectrum}, and as expected theoretically for an internal wave turbulence, there are no significant fluctuations beyond the Brunt-V\"ais\"al\"a frequency $N$. 
Below this frequency the excited modes are homogeneously distributed down to frequencies which are an order of magnitude smaller than both $N$ and $\gamma$ (see figure \ref{fig:mean_temporal_spectrum} left).
These frequencies correspond to the lower branch (\textit{i.e.} small $\xi$ ) of the dispersion relation observed in figures \ref{fig:reldisp_Revariable} and \ref{fig:reldisp_Nvariable}.
When $N$ is increased, \textit{i.e.} when a scale separation appears between the forcing frequency $\gamma$ and $N$, the energy contained in the higher frequencies ($N \sin \xi > 1$) follows a $\omega^{-2}$ power law. 
Such a trend is reminiscent of oceanographic measurements of the velocity which display a similar $\omega^{-2}$ power law in the range of frequencies above the tidal forcing and which is interpreted as a signature of internal wave turbulence \citep{garrett_space-time_1972,garrett_space-time_1975,garrett_internal_1979,levine_modification_2002}. 
To provide a definitive comparison, it would be necessary to increase $N$ while keeping a turbulent saturation, which requires large computational time as the Reynolds number must also be increased. 

As a conclusion, the tidally-driven parametric instability of internal waves saturates in a state reminiscent of ``internal wave turbulence". 
The sustained, broadband frequency and small-scale saturation flow is composed of non-linearly interacting internal waves, although the non-linearity is weak compared to the effects of the background stratification.
\subsection{Anisotropy and decoupling}

\begin{figure}
\centering
\includegraphics[width=0.47\linewidth]{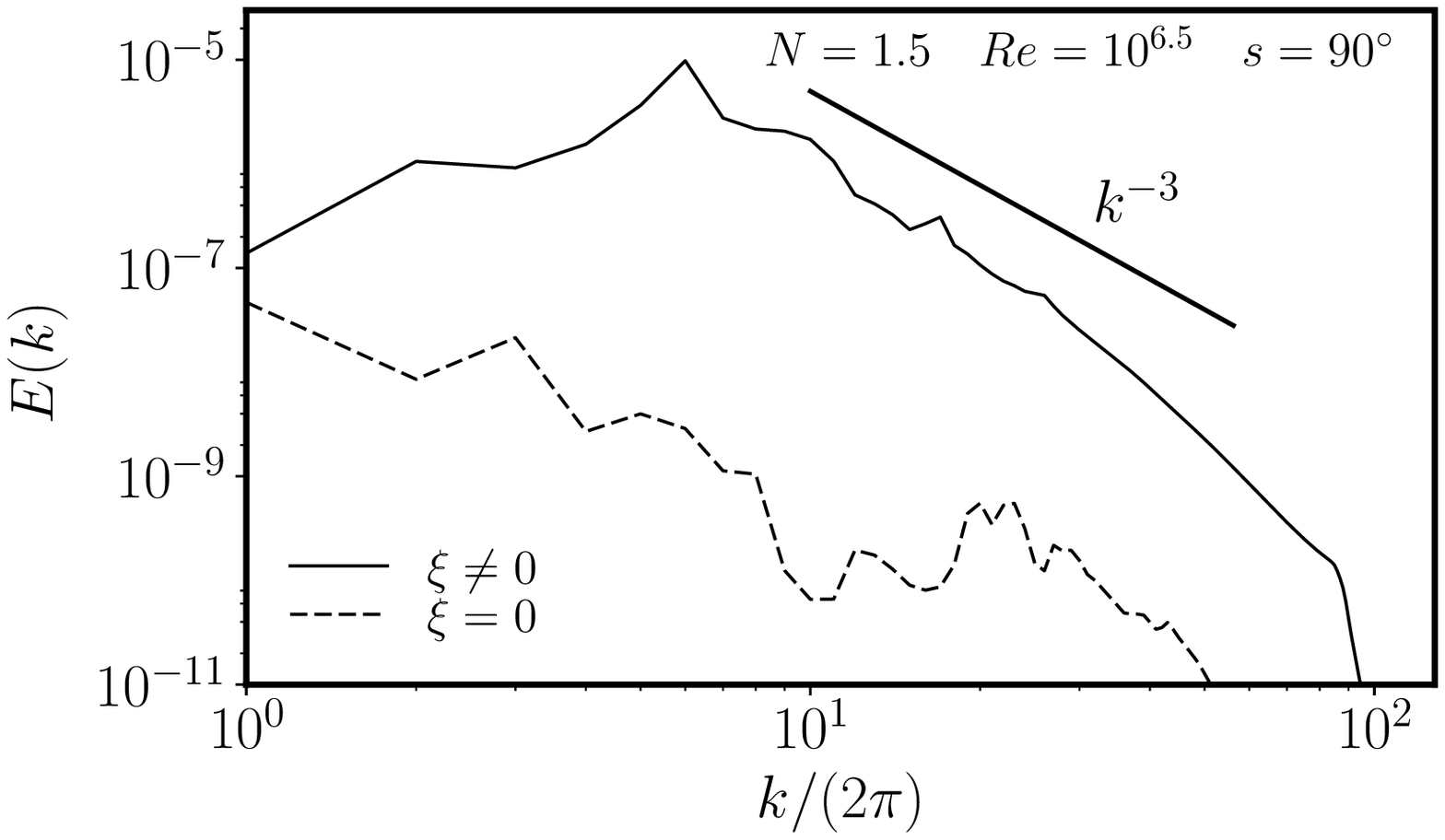}
\includegraphics[width=0.47\linewidth]{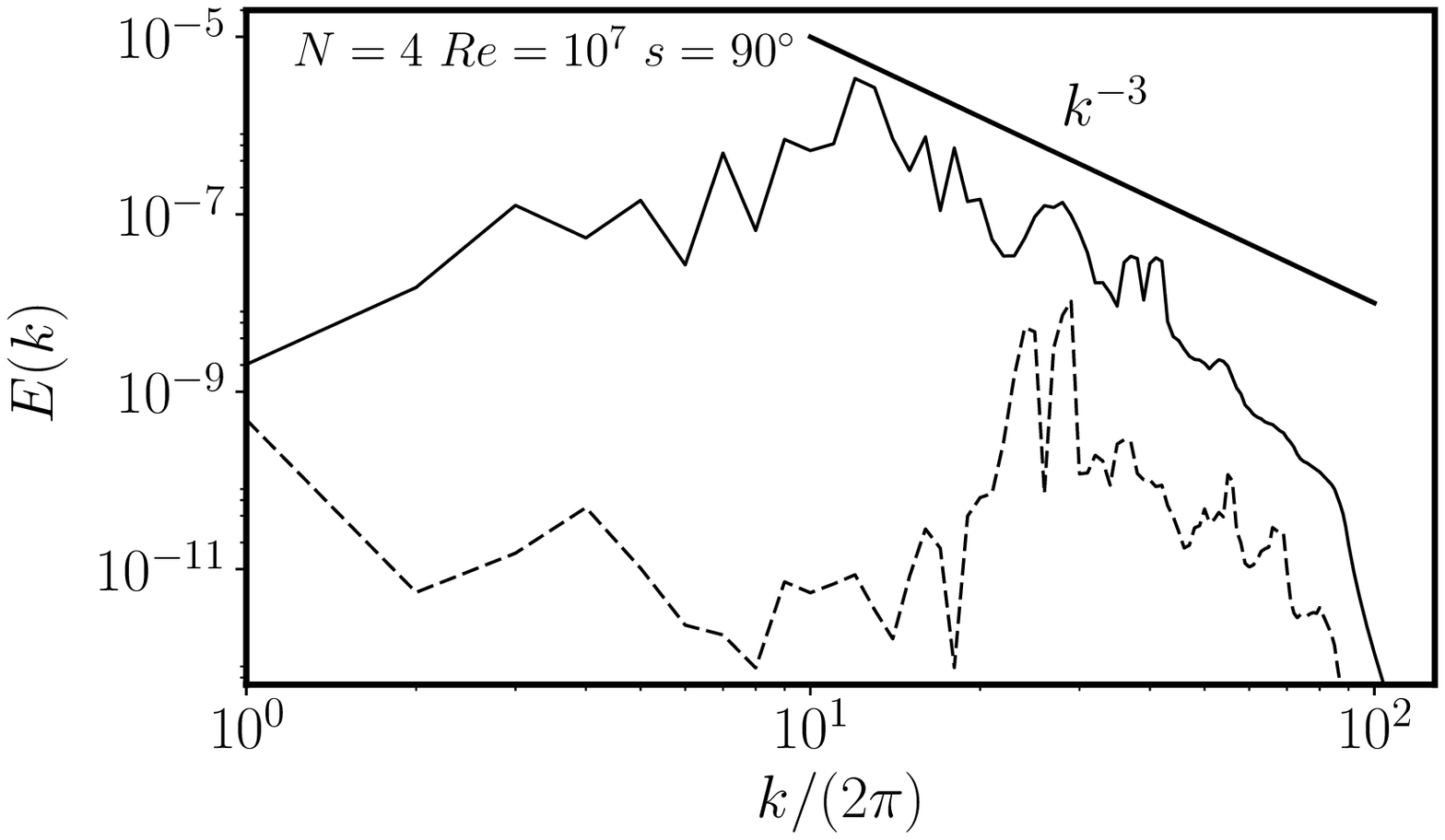}
\caption{
Kinetic energy spectrum of the layerwise slow modes defined by $\xi = 0$ and the rest of the flow for \textbf{left:} $N= 1.5$, $Re= 10^{6.5}$ and \textbf{right:} $N= 4$, $Re= 10^{7}$ at $s = 90^\circ$. 
 }
\label{fig:spectrum_geostrophic}
\end{figure}
\begin{figure}
\centering
\includegraphics[width=0.47\linewidth]{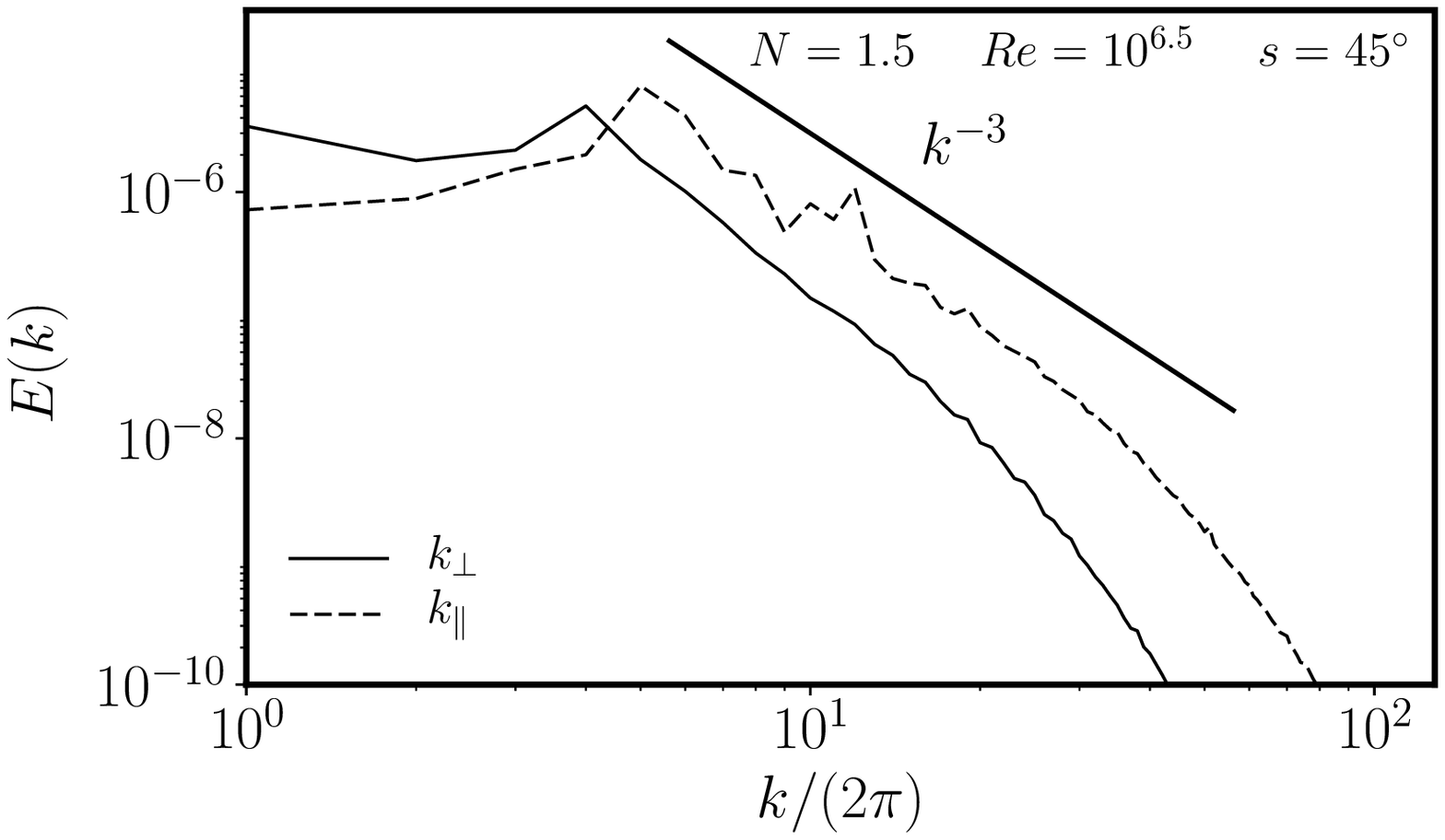}
\includegraphics[width=0.47\linewidth]{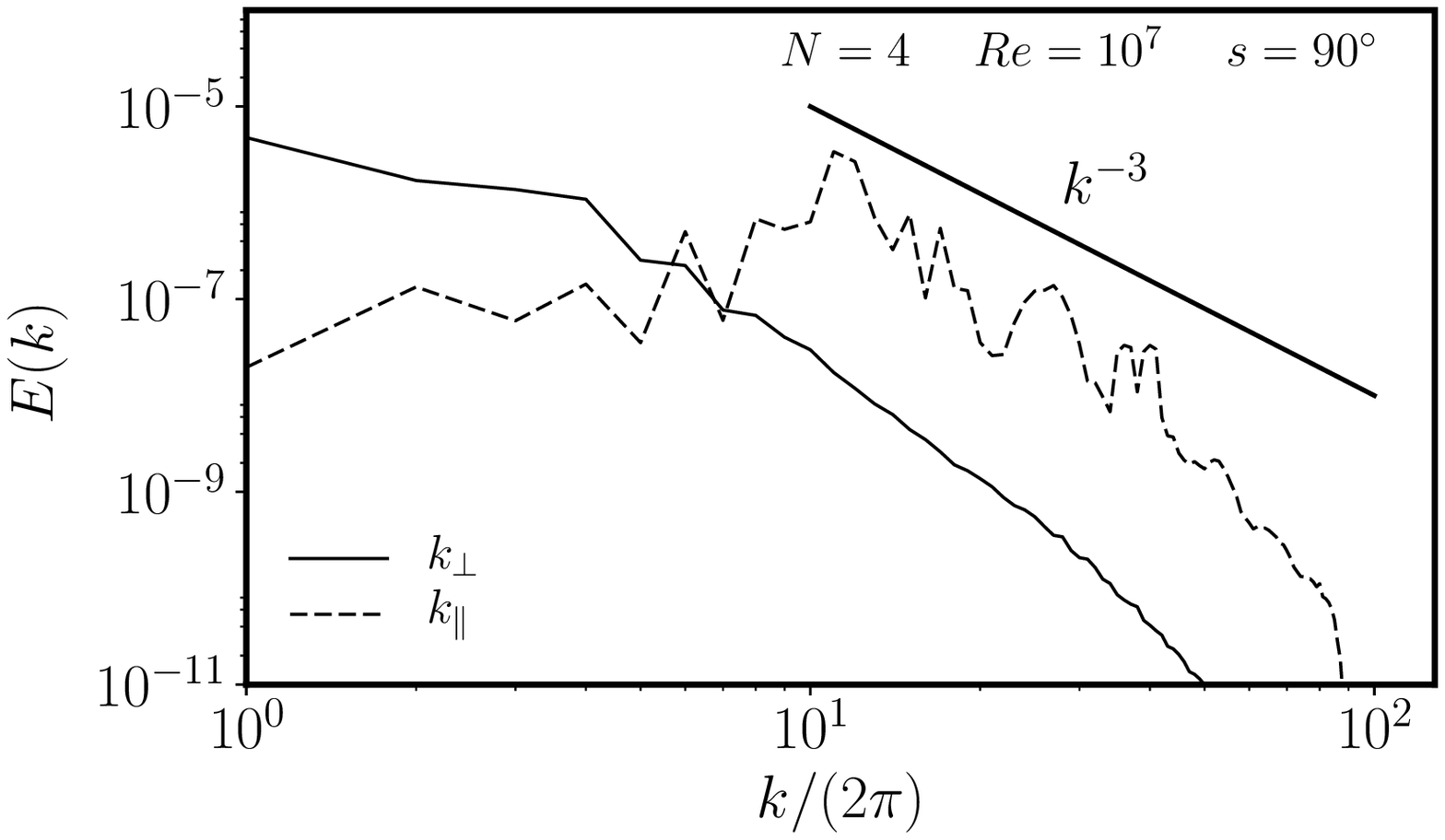}
\caption{
Anisotropic velocity power spectra as functions of $k_\perp$ and $k_\parallel$ for \textbf{left:} $N = 1.5$, $Re = 10^{6.5}$ and $s= 45^\circ$ and \textbf{right:} $N = 4$, $Re = 10^{7}$ and $s= 90^\circ$.  
On each plot, the black solid line materialises a $k^{-3}$ power law for comparison.}
\label{fig:anisotropic_spectra}
\end{figure}

Stratified turbulence has often been studied in the perspective of the emergence of layerwise, low-frequency structures, leading to a strong anisotropy and decoupling between horizontal and vertical variations. 
This paragraph aims at comparing the internal wave turbulence previously identified to the classical theories of stratified turbulence in the high buoyancy Reynolds number regime developed in particular by \cite{billant_self-similarity_2001}, \cite{lindborg_energy_2006} and \cite{brethouwer_scaling_2007}.
First, figures \ref{fig:reldisp_Revariable} and \ref{fig:reldisp_Nvariable} do not indicate any energy accumulation in the layerwise structures around $\xi = 0$ and $\omega = 0$. 
To support this assertion, we show in figure \ref{fig:spectrum_geostrophic} the kinetic energy spectrum for the layerwise modes at $\xi = 0$ and for the rest of the flow. 
As in the two cases the stratification is along the $x$ axis, the  layerwise modes are easily identified in the spectral space as their wave number $\boldsymbol{k}$ is such that $k_z,k_y = 0$. 
As it can be noticed, at all scales, the slow modes are subdominant. 
This could be confusing as layers perpendicular to the stratification can be noticed in figure \ref{fig:snapshot_buy_vorticity}, they however do not correspond to slow modes since they are not exactly invariant along the axes perpendicular to the stratification. 
Although layerwise structures are ubiquitous in stratified turbulence excited by a random forcing or by large scale vortices, forcing waves leads to a completely different state with low energy transfers towards those particular modes. 
The present picture is reversed in the close context of rotating turbulence excited by inertial waves: geostrophic vortices happen to grow up to taking over the whole dynamics in absence of specific dissipative process and to strongly back-react on wave propagation \citep{barker_non-linear_2013,lereun_2017}.
This could be linked to the fundamental mathematical difference between layerwise modes in stratified turbulence and geostrophic vortices in rotating turbulence \citep{cambon_turbulence_2001}, which leads in particular to the absence of inverse cascade in the purely stratified case \citep{marino_inverse_2013,herbert_waves_2016}.

\begin{figure}
\centering
\includegraphics[width=0.5\linewidth]{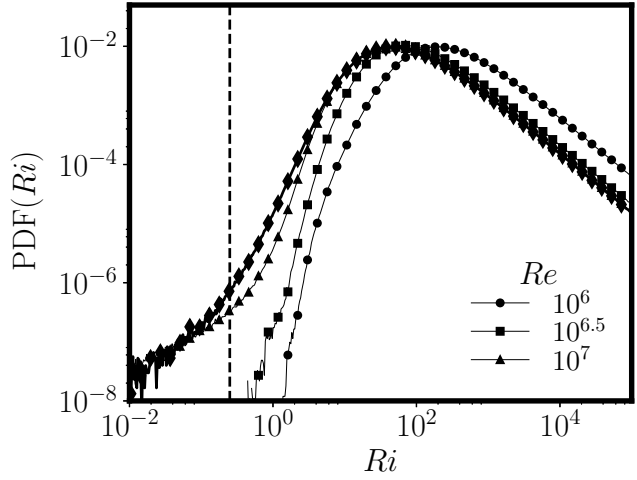}
\caption{Probability density function of the Richardson number for $s= 45^\circ$ and $N = 1.5$  with $Re \in \left\lbrace 10^{6},10^{6.5},10^{7} \right\rbrace$. PDFs are computed from snapshots of the buoyancy and velocity fields, the result presented here is the ensemble average of all the PDFs computed in the saturation phase. The number of samples is usually between 10 and 20. }
\label{fig:richardson}
\end{figure}

This result regarding slow modes has strong implications in the anisotropy of the turbulent flow. 
In the classical theory of stratified turbulence, the velocity power spectrum is an anisotropic function of $k_\perp = k \sin \xi $ and $k_\parallel = k \cos \xi$. 
The velocity power spectrum integrated over $k_\parallel$, \textit{i.e.} along the stratification axis, $E(k_\perp)$, follows a Kolmogorov-like power law $k_\perp^{-5/3}$, while conversely $E(k_\parallel) \propto k_\parallel^{-3}$ \citep{lindborg_energy_2006,brethouwer_scaling_2007}.
However, in the case of the internal wave turbulence presented above, both spectra $E(k_\perp)$ and $E(k_\parallel)$ follow the same power law close to $k_{\parallel,\perp}^{-3}$, as it can be seen in figure \ref{fig:anisotropic_spectra}.
The situation is even reversed as $E(k_\perp)$ is slightly below $E(k_\parallel)$ at large $k$ while it is expected to be dominant in classical stratified turbulence. 
This result shows that there is no decoupling between the horizontal and vertical variations, as observed in the frequently studied high buoyancy Reynolds number regime, which is coherent with the fact that the turbulent  state considered here is a superposition of many internal waves propagating in multiple directions in a quasi-isotropic manner.
The spectra displayed in figure \ref{fig:anisotropic_spectra} suggest that the gradients in the direction perpendicular to the stratification are less steep than they should be if the turbulence was due to shear instability between layerwise modes. 
To investigate whether shear instabilities are possible in the saturated flow, we compute the local Richardson number defined as: 
\begin{equation}
\label{eq:richardson_number}
Ri(\boldsymbol{x},t) = \displaystyle \frac{ N^2 \left( 1 + \displaystyle \frac{\mathrm{d} \vartheta}{\mathrm{d} z_s} (\boldsymbol{x} ) \right) ^2}{\left( \displaystyle \frac{\mathrm{d} \boldsymbol{v}_\perp }{\mathrm{d} z_s} (\boldsymbol{x}) \right)^2}.
\end{equation} 
where $z_s$ is a linear coordinate along the stratification axis and $\boldsymbol{v}_\perp$ is the velocity component perpendicular to the stratification direction. 
It compares the local Brunt-V\"ais\"al\"a frequency, including buoyancy fluctuations, with the shearing rate along the stratification direction. 
Linear stability analysis indicates that a sheared stratified flow is unstable when $Ri < Ri_c = 1/4$. 
Following \cite{brethouwer_scaling_2007}, we compute for several Reynolds numbers, at $N= 1.5$ and $s= 45^\circ$, the PDFs of the local Richardson number. 
As shown in figure \ref{fig:richardson}, for the two lower input  Reynolds numbers, there is no event likely to create shear instabilities.
The buoyancy Reynolds number $\mathcal{R}$ being smaller than one, this is coherent with the remark  of \cite{brethouwer_scaling_2007} that in the low buoyancy Reynolds number regime, there should be no disturbances of Kelvin-Helmholtz type in the flow. 
The picture seems to change at the highest Reynolds number ($10^7$), which corresponds to our most extreme simulation where rare events with $Ri <1/4$ are observed. 
The appearance of rare unstable events could be reminiscent of a transition towards a high buoyancy Reynolds number regime, which is further discussed in the concluding section of the present paper. 
Still we conclude that in the regime we explore in the present paper, the internal wave turbulence is mostly stable to shear instabilities and is unable to drive strong overturning events. 
\subsection{Mixing}
\label{mixing}
\begin{figure}
\centering
\includegraphics[width=0.90\linewidth]{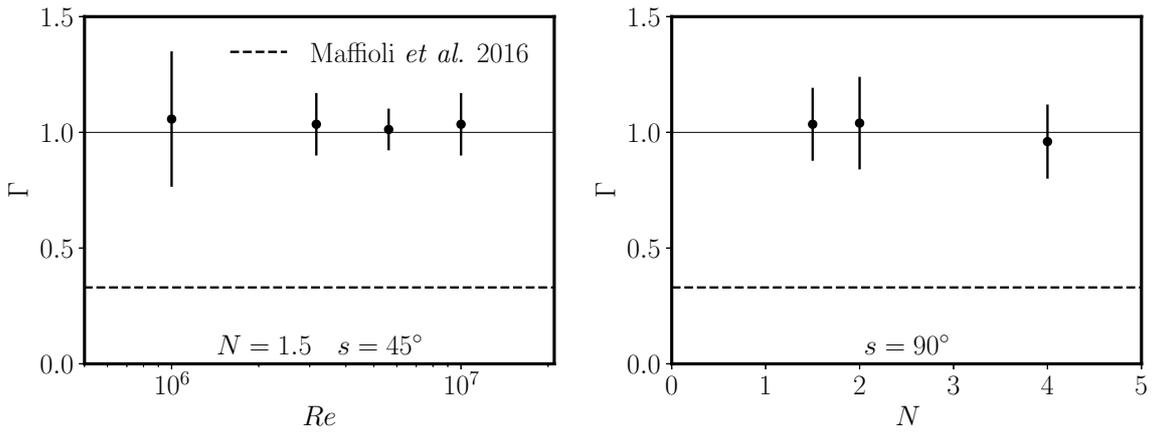}
\caption{
\textbf{Left:} mixing coefficient $\Gamma$ as defined by \citep{maffioli_mixing_2016} as a function of the input Reynold number.
The limit value in the small Froude regime obtained by \cite{maffioli_mixing_2016} is given as a reference. 
\textbf{Right:} evolution of the mixing coefficient with $N$ for $s = 90^\circ$ and $Re = 10^{6.5}$ except at $N = 4$ where $Re = 10^7$.
}
\label{fig:mixing_efficiency}
\end{figure}

At this stage, we know that tides are able to amplify buoyancy perturbations over a background stratification, and that this amplification saturates into an internal wave turbulence.
We would like to quantify then how this turbulent state mixes the buoyancy perturbations, \textit{i.e.} how it irreversibly converts potential energy perturbations into background potential energy \citep{peltier_mixing_2003}. 
Following the work of \cite{lindborg_vertical_2008}, \cite{salehipour_diapycnal_2015} and \cite{maffioli_mixing_2016}, we propose to quantify the mixing via a coefficient $\Gamma$ defined as:
\begin{equation}
\label{eq:mixing_coefficient}
\Gamma ~=~ \frac{\varepsilon_p}{\varepsilon_{\rm{k}}}
\end{equation} 
where $\varepsilon_p = -  N^2 (Re ~ Pr )^{-1} \left\langle (\boldsymbol{\nabla} \vartheta)^2 \right\rangle$ quantifies the diffusion of the buoyancy perturbations and $\varepsilon_{\rm{k}}$ is the kinetic energy dissipation defined earlier. 
This coefficient $\Gamma$ was originally introduced to evaluate how turbulence induces an effective diapycnal diffusivity $K = \Gamma \varepsilon_{\rm{k}}/N^2$ \citep{osborn_estimates_1980,salehipour_new_2016}.
The expression given for $\varepsilon_p$ can be retrieved by considering that it is a potential energy dissipation \citep{lindborg_vertical_2008}. 
In the derivation of the energy equations from (\ref{eq:NS_buoy_adim}), the buoyancy equation must be multiplied by $N^2 \vartheta$ to obtain the same energy transfer from velocity to buoyancy, which finally yields to the definition given earlier to $\varepsilon_p$.

In their study, \cite{maffioli_mixing_2016} found that, forcing a turbulence in a stratified fluid with vortices aligned with stratification, at low Froude and high buoyancy Reynolds numbers the mixing coefficient $\Gamma$ converges towards $0.33$. 
In figure \ref{fig:mixing_efficiency} (left), we display the mixing coefficient $\Gamma$ at $N = 1.5$ as a function of the input Reynolds number $Re$ with the limit value found by \cite{maffioli_mixing_2016} given as a reference.
Despite $Re$ is increased from $10^6$ up to $10^7$ 
 we do not observe any variation of the mixing coefficient. 
Instead, $\Gamma$ remains constant around $1$, well above the limit reference value.
The evolution of the mixing coefficient $\Gamma$ with the Brunt-V\"ais\"al\"a frequency $N$ is also computed for $s = 90^\circ$. 
As it can be noticed in figure \ref{fig:mixing_efficiency} (right), again, $\Gamma$ remains constant and around $1$. 
This result can be inferred from a very simple model assuming the flow is only a superposition of low amplitude internal waves with weak non-linear interactions. 
A single wave of frequency $\omega$ and wave vector $\boldsymbol{k}$, $\left\lbrace \boldsymbol{u} , \vartheta, \Pi \right\rbrace = \left\lbrace \boldsymbol{u}_{\boldsymbol{k}0} , \vartheta_{\boldsymbol{k}0}, \Pi_{\boldsymbol{k}0} \right\rbrace
e^{i (\boldsymbol{k} \cdot \boldsymbol{x} -\omega t)}
$ with $\omega^2 = N^2 \sin \xi$, must obey the following linear inviscid set of equations:
\begin{equation}
\label{eq:lin_inv_WT}
\left\lbrace
\begin{array}{rl}
 \partial_t \boldsymbol{u} & = - \boldsymbol{\nabla} \Pi +  N^2 \vartheta \boldsymbol{e}_s \\
 \partial_t \vartheta  &= -\boldsymbol{e}_s \cdot \boldsymbol{u} 
\end{array}
\right.
\Rightarrow
\left\lbrace
\begin{array}{rl}
 - i \omega ~\boldsymbol{u}_{\boldsymbol{k}0} & = - \boldsymbol{k} \Pi_{\boldsymbol{k}0} +  N^2 \vartheta_{\boldsymbol{k}0} ~ \boldsymbol{e}_s \\
 - i \omega ~\vartheta_{\boldsymbol{k}0}  &= -\boldsymbol{e}_s \cdot \boldsymbol{u}_{\boldsymbol{k}0}
\end{array}
\right.
\end{equation}
where $\boldsymbol{e}_s$ is the stratification direction unit vector. 
We wish then to compute the volume averaged dissipation associated with the wave field, which is merely the sum of each single wave dissipation. 
For one wave only, $\left\langle \boldsymbol{u}^* \cdot \boldsymbol{\nabla}^2 \boldsymbol{u} \right\rangle = - k^2 ~\vert\boldsymbol{u}_{\boldsymbol{k}0} \vert^2 $ and $\left\langle \vartheta^*  \boldsymbol{\nabla}^2 \vartheta \right\rangle = - k^2 ~\vert \vartheta_{\boldsymbol{k}0} \vert^2 $.
To compute $\Gamma$, we need to relate $\vert\boldsymbol{u}_{\boldsymbol{k}0} \vert^2$ to  $\vert \vartheta_{\boldsymbol{k}0} \vert^2$ which can be done for instance applying $\boldsymbol{k} \times ( \boldsymbol{k} \times \cdot )$ to the velocity equation in (\ref{eq:lin_inv_WT}).
We then obtain the exact balance $\vert\boldsymbol{u}_{\boldsymbol{k}0} \vert^2 = N^2 \vert \vartheta_{\boldsymbol{k}0} \vert^2$ (which does not apply at $\omega = 0$).
Thus we find with the following simple scaling for the mixing coefficient:
\begin{equation}
\Gamma = \frac{1}{Pr}
\end{equation}
where $Pr$ is the Prandtl or Schmidt number. 
We retrieve for our simulations at $Pr= 1$ that $\Gamma = 1$. 
The numerical result $\Gamma = 1$ should therefore be regarded as an additional signature of internal wave turbulence.

To conclude, internal wave turbulence offers a picture completely different from the classical stratified turbulence at high buoyancy Reynolds number. 
The flow being a superposition of low to moderate amplitude waves, the mixing coefficient is different compared to a situation where the most energetic structures are the non-propagative layerwise modes.
Note that although the mixing coefficient is increased, the consequent turbulent diapycnal diffusivity should still be lower than what is measured in high buoyancy Reynolds number regime, essentially because the forcing introduced here and the associated dissipation rates are small.
\section{Conclusion}
Throughout this paper, we have shown with an idealised local Lagrangian model that tidal flows are able to drive bulk turbulence in stratified planetary fluid layers. 
This turbulence is driven by the parametric subharmonic resonance of unstable internal waves. 
The latter continuously feeds a cascade of daughter waves to create a flow which bears signatures of internal wave turbulence, in particular the focalisation of the energy along the dispersion relation of internal waves. 
Such a turbulent flow has already been characterised in an experimental setup designed to mimic the effects of tides on a particular topography \citep{brouzet_energy_2016}.
We claim from our results that internal wave turbulence can take place homogeneously in a whole fluid layer undergoing tidal distortion, provided that the latter's amplitude is large enough to overcome dissipation and that the Brunt-V\"ais\"al\"a frequency $N$ is larger than the tidal frequency $\gamma$. 
In addition, our local approach provides an efficient way to numerically investigate the detailed and possibly universal properties of weakly forced internal wave turbulence in low dissipation regimes.
Despite recent experimental \citep{brouzet_energy_2016} and theoretical \citep{gamba_2017} advances, this particular state of stratified turbulence remains challenging and difficult to be compared to the classical theory of wave turbulence \citep{zakharov_kolmogorov_2012,nazarenko_wave_2011}. 
As in rotating turbulence, this is essentially due to the anisotropy of the dispersion relation (see for instance the discussion in \cite{brouzet_internal_2016_thesis}), the role played by near-resonant interactions and the non-linear interaction with non-propagative modes  \citep{cambon_turbulence_2001,galtier_weak_2003,
smith_near_2005,
bellet_wave_2006,
scott_wave_2014,
gelash_complete_2017,
gamba_2017}.
Although our model is introduced in a targeted geophysical context, it could be used to test universal internal wave turbulence models or closure. 

Future work will strive to introduce rotation, which is also a key ingredient to planetary fluid dynamics. 
In particular, in the limit where buoyancy effects still dominate over the Coriolis force, it should be interesting to study the consequence of its introduction on the filling of the dispersion relation and the subsequent low-frequency cut-off.
As it can be noticed in figure \ref{fig:mean_temporal_spectrum}, for $N=4$, there is an energy accumulation at frequencies $\omega \in \left[ 0.1, 1 \right]$. 
If the rotation rate were to be in this range, would the energy accumulate in the lowest frequency modes, \textit{i.e.} the layerwise structures which were never observed to develop in our simulations? 
If so, tidal flows would convey energy into modes which could then undergo shear instabilities and therefore drive more intense turbulence with enhanced dissipation rate and mixing. 
Whether this turbulence is sustained or lead to the temporary inhibition of the instability which feeds it remains to be seen. 
Moreover, it would be interesting to investigate the persistence of the results found here in the regime of high or low Prandtl (or Schmidt) number, which are both relevant to geophysical fluid dynamics. 
Note though that it is already known from the experiments of \cite{brouzet_energy_2016} that internal wave turbulence can be excited in salted water, \textit{i.e.} at high Schmidt number. 
As linear internal waves are characterised by energy equipartition (see paragraph \ref{mixing}), we should expect that the resonant energy transfer towards small scale is inhibited as soon as either viscosity or diffusion balances non-linear advective transfer. 
What happens to the non-dissipated quantity and how it interacts with the larger scale waves beyond this cut-off remains an open question.
The Prandtl (or Schmidt) number should not play any significant role in the large scale behaviour of the flow. 
\begin{figure}
\centering
\includegraphics[width=0.5\linewidth]{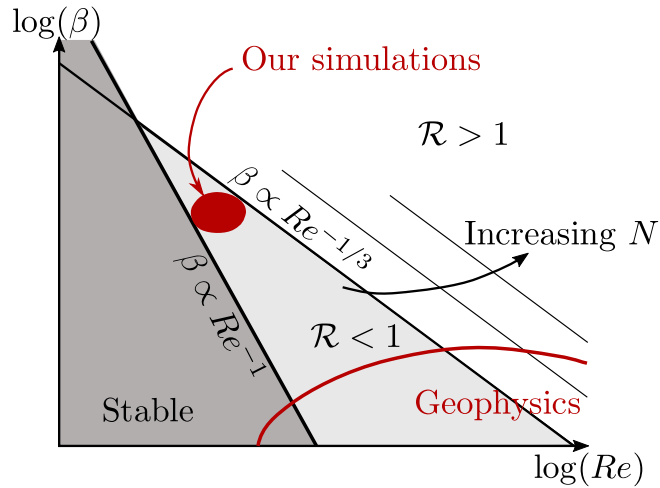}
\caption{Schematic relative distribution of the low and high buoyancy Reynolds number $\mathcal{R}$ regimes as a function of the input Reynolds number $Re$ and the ellipticity of the deformation $\beta$.
The area covered by geophysical regimes and our simulations is indicated in particular to highlight the fact that the low $\mathcal{R}$ regime is also relevant to planetary layers in the case of bulk wave turbulence excited by tides. }
\label{fig:regime_diagram}
\end{figure} 
Lastly, the type of turbulence resulting from the saturation of the tidally-driven instability occurs in a regime of high Reynolds, low Froude and low buoyancy Reynolds ($\mathcal{R}$) numbers, which leads to a completely different picture compared to the high $\mathcal{R}$ regime frequently studied and branded as the regime relevant to geophysical flows. 
In particular, the mixing coefficient is increased in the regime we describe, and is coherent with the scaling $\Gamma = 1/ Pr$ that we have derived theoretically assuming the flow is a superposition of linear internal waves only. 
This result is an additional signature of wave turbulence.
However, this enhanced mixing coefficient may not result in an increase in the turbulent diapycnal diffusivity as the forcing and the dissipation are small. 

A regime of high buoyancy Reynolds number turbulence excited by the parametric instability introduced here is possible in the very high Reynolds number limit, but could not be investigated because it is highly demanding in resolution and computational time, or it requires an increase of the ellipticity $\beta$ to unrealistic values.
As seen earlier, events with Richardson number under 1/4 are measured in the most extreme simulation where the dissipation is so small that the buoyancy Reynolds number reaches $\sim 5$.
It would be interesting to see whether at high $\mathcal{R}$ wave turbulence can drive strong over-turning events or not, and how it would impact the mixing coefficient and the turbulent diapycnal viscosity. 
Still we claim that both regimes should be considered as relevant to geophysical applications due to the specificity of our forcing mechanism favouring weak wave interactions. 
Indeed, the buoyancy Reynolds number $\mathcal{R}$ can be expanded as
\begin{equation}
\mathcal{R} = \frac{u_{\mathrm{rms}}^3 Re}{\lambda_{\mathrm{res}} N^2}.
\end{equation}
Assuming that the saturation results from the balance between the forcing term $\mathsfbi{A}(t) \boldsymbol{u} \sim \beta u_{\mathrm{rms}}$ and the non-linear term  $\boldsymbol{u} \cdot \boldsymbol{\nabla } \boldsymbol{u} \sim u_{\mathrm{rms}}^2 / \lambda_{\mathrm{res}}$ leads to $u_{\mathrm{rms}} \sim \beta \lambda_{\mathrm{res}}$.
As a result, the buoyancy Reynolds number goes like: 
\begin{equation}
\mathcal{R} \sim \beta^3 Re \left(\frac{\lambda_{\mathrm{res}}}{N}\right)^{2}.
\end{equation}
The area with high $\mathcal{R}$ lies above a line $\beta \propto Re^{-1/3} ~(N/\lambda_{\mathrm{res}})^{2/3}$.
In addition, the instability grows when the forcing overcomes the volume viscous dissipation, \textit{i.e. } for $\beta \gtrsim (\lambda_{\mathrm{res}}^2 Re)^{-1}$.
As a result, in the $(\beta,Re)$ plane, both regimes are worth considering in the geophysical limit where usually $\beta$ is smaller than $10^{-3}$ and $Re$ is large, as indicated in figure \ref{fig:regime_diagram}.
Note that this discussion is unchanged if we consider the dissipation to be due to solid wall friction, for which the unstable zone lies above the line $\beta \propto Re^{-1/2}$.
Moreover, as indicated in figure \ref{fig:regime_diagram}, the area of small $\mathcal{R}$ is extended as $N$ is increased.  
In future work, it would be interesting to delimit more precisely those two regimes. 
Note that a possible transition could be approached in our most extreme simulation for which $\mathcal{R} \sim 5$. 
Exploring the internal wave turbulence driven at high buoyancy Reynolds number would therefore require increasing the ellipticity and thus the forcing intensity. 
This, we believe, deserves a study of its own. 

Lastly, we believe the results presented here should not change as the ellipticity is lowered provided that $\mathcal{R} < 1$ and the flow is unstable.
In addition, when three waves of frequencies $(\omega_1,\omega_2,\omega_3)$ exchange energy via triadic resonance, the resonance condition on frequency must be satisfied with a tolerance $\mathcal{O}(Fr)$ \textit{i.e} $\omega_1 \pm \omega_2 \pm \omega_3 = \mathcal{O}(Fr)$ due to detuning by larger scales advection (see \cite{smith_near_2005} for a discussion in the analog context of inertial waves in rotating flows). 
As $u_{\mathrm{rms}}$ scales like $ \beta \lambda_{\mathrm{res}}$, decreasing $\beta$ corresponds to decreasing $Fr$ and therefore to more exact resonances. 
The only significant change, we believe, is a thiner focalisation of the energy along the dispersion relation of internal waves. 
\textbf{Acknowledgement:} We acknowledge support from the European Research Council (ERC) under the European Union's Horizon 2020 research and innovation program (grant agreement No. 681835-FLUDYCO-ERC-2015-CoG). We also acknowledge support from IDRIS (Institut du D\'eveloppement et des Ressources en Informatique Scientifique) for computational time on Turing (Projects No. 100508 and 100614) and from the HPC resources of Aix-Marseille Universit\'e (Projects No.15b011 and 16b020) financed by the project Equip@Meso (No. ANR-10-EQPX-29-01) of the program Investissements d'Avenir supervised by the Agence Nationale pour la Recherche.


\end{document}